\documentclass[12pt]{iopart}

\usepackage{iopams}  
\usepackage{graphicx}  

\newcommand{\dd}		{\mathrm{d}}

\begin{document}

\title[Collective Flow of QCD Matter: a Historical Introduction]{Collective Flow of QCD Matter: a Historical Introduction}

\author{Hans Georg Ritter$^{1}$ and Reinhard Stock$^{2,3}$}

\address{
1 Nuclear Science Division, Lawrence Berkeley National Laboratory,\\~~\,Berkeley, Ca., USA
}
\address{
2 Institute of Nuclear Physics, Frankfurt University
}
\address{
3 Frankfurt Institute for Advanced Stuies, Frankfurt University
}
\eads{\mailto{hgritter@lbl.gov}, \mailto{stock@ikf.uni-frankfurt.de}}


\section{Introduction} \label{SectI}

We encounter extended QCD matter in nuclei, neutron stars and supernovae, and in the early universe. Only nuclei permit a study of matter properties in the terrestrial laboratory, and every textbook begins with the relatively tight packing of nucleons. Their average distance in nuclear ground state matter is about 1.8~fm, just about equal to twice the nucleon radius from electron scattering: we are dealing with a (quantum) liquid. From among its
numerous collective excitations we recall, in particular, the so-called
giant monopole excitation mode \cite{1} which was understood as a global compression-expansion oscillation of the entire nuclear density distribution, around the ground state density $\rho_{0}$ of about 0.16 nucleons per~fm$^{3}$. This involves the compressibility (and the sound velocity) of zero temperature nuclear matter and, more generally, the equation of state (EOS). The latter describes the relation between density and pressure, the restoring force of the GMR oscillation mode in a quasi-classical description which showed that the relation of energy to density (oftentimes called EOS in the early literature) was parabolic about $\rho_0$, leading to a linear response system. The monopole excitation mode, observed in medium mass nuclei \cite{1}, turned out to sample only an extremely narrow density interval about $\rho_{0}$. On the other hand, concurrent late 1960's investigations of supernova explosion dynamics \cite{2}, and of neutron star radial density profiles \cite{3} had shown that nuclear matter densities well in excess of 3 $\rho_{0}$ should govern the astrophysical phenomena. Extrapolation of the parabolic GMR EOS, however, resulted in nuclear sound speeds $c^{2}_{s}= \partial p/\partial \rho$ exceeding light velocity, an "acausal solution":
the rise of that EOS was too steep at high densities. This defined a major, new paradigm of theory and experiment: hadronic matter at high density, and its equation of state (the quest for the EOS of partonic matter followed about ten years later).

The EOS represents the major "nontrivial" ingredient in a hydrodynamical description of hadronic matter flow: for "closing" the set of ideal hydrodynamics equations which, otherwise, are derived entirely
from a priori conservation of energy and momentum in 3+1 dimensions.
The Frankfurt theory group headed by W.Greiner presented the first proposal
to compress ground state nuclear matter in central collisions of heavy
nuclei \cite{4}, to densities far beyond 2 $\rho_{0}$, the naive overlap density, and
to analyze the to-be-expected collective hydrodynamic de-excitation flow
modes to determine the underlying EOS at high density.

Is hadronic matter hydrodynamic ("shock flow" \cite{4}) expansion realistic in this environment? Recall a precondition for the validity of a hydro description, the zero mean free path requirement. It turns out that the mean free path length $\lambda = (\sigma\rho)^{-1}$ is about as small as it could be in nuclear matter, as a consequence of the strong interaction: each nucleon interacts with every
other nucleon it comes across, thus lambda
is already as low as about 2~fm at ground state density, equal to the average spacing between nucleons. This is small in comparison to the size of the "fireball" volume for heavy nuclei, and it reduces toward one fermi at the initial high density. The strong interaction cross section is as large as the geometrical transverse area of a hadron, "everybody pushes everybody else": a typical flow condition. Of course, toward final decoupling at the end of the expansive evolution the zero mean free path description becomes ill-satisfied, and one needs to switch to an "afterburner" transport description.

Second, however, the application of an EOS requires, at least local kinetic equilibrium. This does not arise instantaneously, in principle one needs to employ an "initialization period". Equilibrium in these relatively small systems is an issue that one should not discuss with fundamentally minded theoretical colleagues. Of course the A+A collisional volume, of some 6~fm radius, can not reach global uniform density and temperature faster than about 6~fm$/c$. Intuition, as supported by microscopic parton or hadron transport models, suggests that there should occur initial energy density inhomogeneities (see below), as well as surface dilution effects. However this is not the point, flow occurs in local "cells", each containing perhaps up to hundred degrees of freedom that support a local description of pressure and flow velocity fields. In fact, most of the collective flow observables with which we shall deal in this introduction, and in the entire present volume, represent non-isotropic, directed emission patterns, of complex spatial structure, which can be interpreted to reflect non-trivial, non-uniform distributions of pressure. Of course non-isotropic emission will also have non-flow causes, such as finite number fluctuations, Bose-Einstein correlations, and jets (at high energy). These contributions require non-trivial background corrections in the data analysis.

A key example of non-isotropic emission is the so-called elliptic flow \cite{5}; we shall turn to it in
more detail later but want to underpin here what it has to tell us about
the equilibrium issue. In fact it represents the most striking manifestation of the primordial dynamical conditions, directly after interpenetration of the projectile-target density distributions. It reflects a snapshot of the geometrical arrangements prevailing, at top RHIC and LHC energies, for sub-fractions of a~fm$/c$ only, but this appears enough to become imprinted into a prominent flow mode. A striking manifestation for an early onset of hydrodynamic evolution, revealing an initialization time of a mere 0.5~fm$/c$, elapsed after initial interpenetration. This signal thus shows sensitivity to the primordial partonic EOS (as well as to other collective properties) of the QCD quark-gluon plasma state. We finally conclude that local equilibrium appears to be attained after an extremely short initialization time: still a challenge to theory. It remains to say that a state of the art hydrodynamic description of the collisional evolution requires, both, a separate initialization model, and a decoupling model for the post-equilibrium stage of final hadronic expansion. Of course, this is a recent insight. Let us first turn to the discovery of flow.

Directed hydrodynamical flow was discovered at the LBL Bevalac in the early 1980's, by the GSI-LBL Plastic Ball Collaboration \cite{6,7,8}. This came about by a unique coincidence of several critical innovations. First, the Bevalac provided for the first genuinely relativistic heavy nuclear projectiles, from $^{40}$Ca via $^{93}$Nb to $^{196}$Au, at energies ranging from 0.2 to 1.8~GeV per projectile nucleon ($1.46<\sqrt{s}<2.27$ GeV). Second, at such energies the charged particle multiplicities in semi-central and central collision events rose toward hundred. At such high multiplicities the individual events become "self-analyzing" with respect to "giant" collective matter expansion modes, thus allowing for a statistically significant analysis: the advent of event-by-event physics. An example: collective sideward directed proton flow (today called $v_{1}$) occurs in the reaction plane spanned by the directions of the beam, and the target-projectile impact vector. In an ensemble of collision events this plane rotates randomly about the beam axis. With high event-wise proton multiplicities one can approximately reconstruct the principal axis of the multiparticle momentum space flow tensor, event by event; it points in the direction of the impact
vector. All events are then rotated into a common reaction plane. The collisional geometry thus gets (approximately) fixed in a new ensemble, and
flow observables that are related to the reaction plane can be determined
with high ensemble statistics.

As a further innovative side effect, the high charged particle multiplicities
allowed for selection of successive multiplicity windows, the corresponding sub-ensembles characterized by successive (but overlapping) domains of impact parameter. Thus one can also, more broadly, select for central, semi-peripheral and peripheral collisional geometries. The initial geometry
fixes the initial state of the dynamic evolution, essential for the
understanding of directed flow observables. We note that all these innovations at Bevalac times are standard ingredients of A+A analysis until today.

Finally: the crucial development of detector systems permitting a "charged particle exclusive" analysis, i.e. detection of multiparticle flow, took place at the LBL Bevalac. Both the Plastic Ball \cite{9} and the Streamer Chamber \cite{10} detector systems were "4$\pi$ acceptance" devices: to record, ideally, the entire charged particle production from an A+A fireball. Recall that at the low Bevalac energies the total final momentum space is still relatively narrow, all produced particles confined to a rapidity interval $|y_{CM}|< 1$. A suitable detector system could thus record almost the entire event, albeit with certain losses at low $p_{T}$, but highly satisfactory at mid-rapidity which is most decisive for flow analysis.

\begin{figure}[ht]
  \centering
  \includegraphics[width=0.5\textwidth]{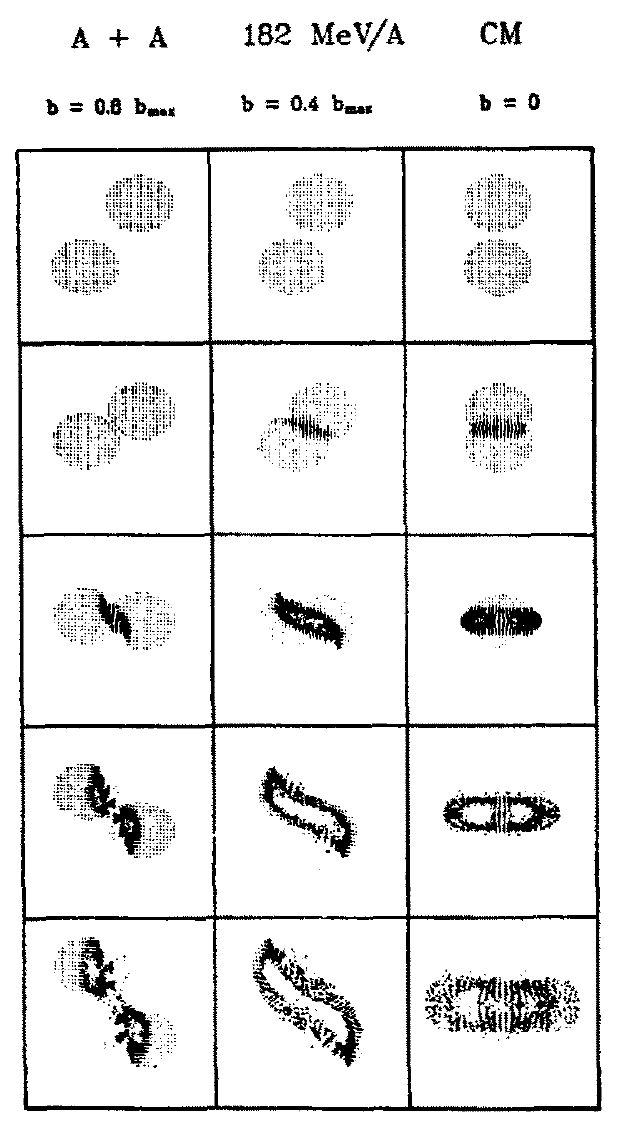}
  \caption{
    Early prediction of the hydrodynamic model \cite{11}, illustrating the temporal evolution in A+A collisions for central, semi-central and peripheral impact geometries. Radial flow results for b=0 whereas the b=0.4~b$_{max}$ panel exemplifies a clear prediction of directed sideward flow.
    \label{fig1}
  }
\end{figure}

\begin{figure}[ht]
  \centering
  \includegraphics[width=0.5\textwidth]{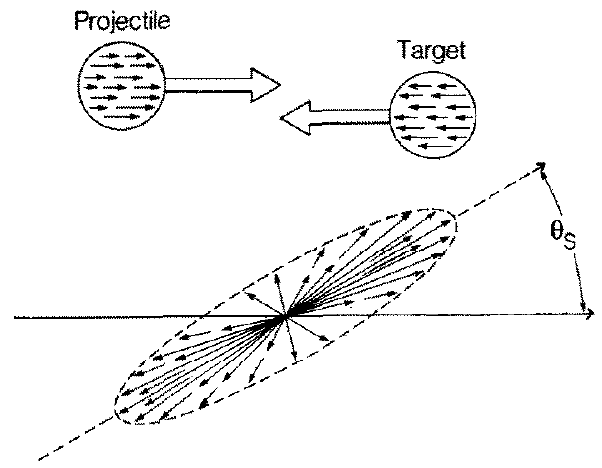}
  \caption{
    Schematic illustration of sphericity analysis event by event. Diagonalization of the momentum tensor in the cm frame for all charged particles leads to a triaxial ellipsoidal shape. The angle of the principal axis relative to the beam direction is the flow angle. Together with the beam direction it defines the reaction plane.
    \label{fig2}
  }
\end{figure}

\begin{figure}[ht]
  \centering
  \includegraphics[width=0.6\textwidth]{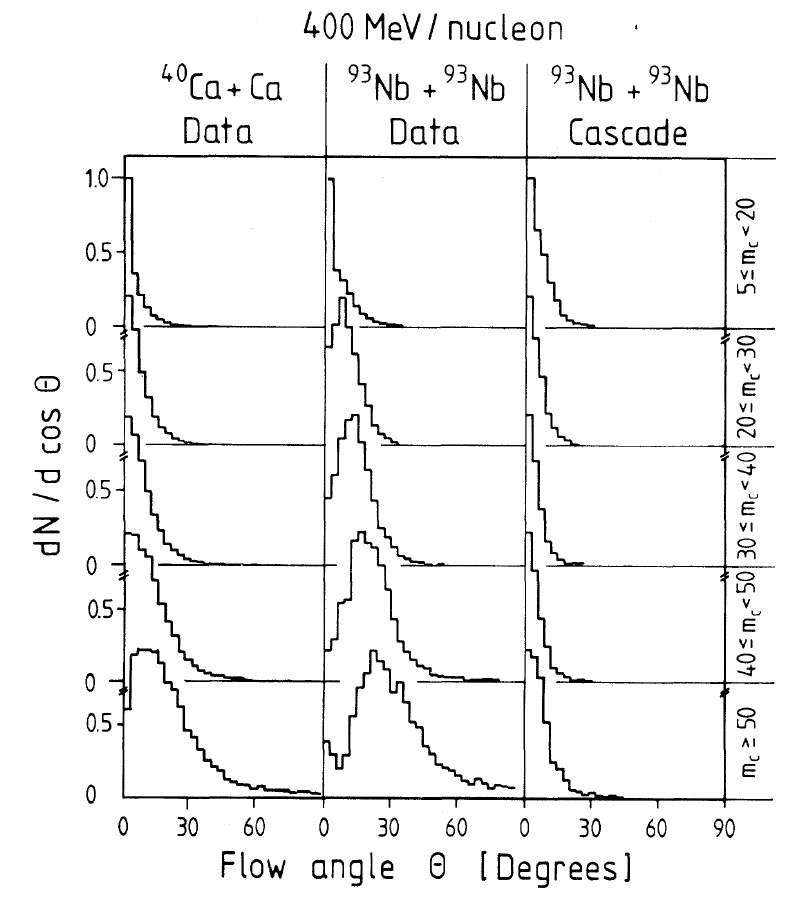}
  \caption{
    The discovery of directed sideward flow \cite{6} at the Bevalac. A clear signal of sideward flow is seen in semi-central Nb+Nb collisions at 0.4 GeV/nucl. It is absent in corresponding results from a hadron gas transport model \cite{12}.
    \label{fig3}
  }
\end{figure}

We illustrate these early accomplishments in Figures \ref{fig1} to \ref{fig3}. An example for the predictions of the hydrodynamic model \cite{11} is shown in \Fref{fig1}. The semi-peripheral event class (middle column) exhibits a spatial evolution leading to an ideal image of sidewards directed matter flow; "ideal" because the reaction plane is exactly known in the model.

\Fref{fig2} illustrates the analysis method that employed the sphericity tensor in momentum space after event by event determination of the reaction plane, the principal axis pointing out the flow angle in plane. \Fref{fig3} recollects the discovery of directed sideward flow, by the Plastic Ball Collaboration \cite{6} in 1984. It shows the flow angle histograms in $^{40}$Ca+$^{40}$Ca and $^{93}$Nb+$^{93}$Nb collisions, for multiplicity bins ascending toward semi-central collisions. Sideward flow is clearly visible in the Nb data whereas the results for the lighter Ca projectile show a marginal signal only. A flow signal is shown to be absent in Nb+Nb calculations with a microscopic hadron transport model \cite{12} which represents the dynamical evolution of a hadron-resonance-"gas".

The 1984 conclusion: high density compressed hadronic matter behaves like a liquid, and, moreover, these data should allow for an extraction of the equation of state at high density. However the high temperature is a principal initial obstacle
to determining the EOS of cold neutron stars. It took decades of theoretical work \cite{13} to extrapolate to zero temperature.
Quite on the contrary, the big bang was at high temperature, and the initial temperatures of 200 to 350 MeV that are encountered in the initial hydro flow periods at RHIC and LHC energies \cite{14} do, in fact, represent directly the big bang conditions during the first microseconds era. These are governed by the QCD quark-gluon plasma state: the principal goal of all contemporary nucleus-nucleus collision physics.

A final look at \Fref{fig3} recalls a further lesson learned at the Bevalac: the need for heavy nuclear projectiles such as $^{196}$Au, $^{208}$Pb, or even $^{238}$U, and the benefits of studying collisions of equal mass nuclei. Both these preferences follow from simple intuition. Compare an ideal fireball formed in central collisions of $^{20}$Ne+$^{20}$Ne, and of $^{208}$Pb+$^{208}$Pb. At a hypothetical initial energy density of, say, 3~GeV$/$fm$^{3}$ (i.e. well above QCD deconfinement) the corresponding fireball radii will be about 3, and 7~fm, respectively. Assuming, for simplicity, a uniform surface expansion speed of $\beta=0.7$, the density will fall down to 1 GeV$/$fm$^3$ (approximately corresponding to QCD confinement) within 1.9 and 4.3~fm$/c$, respectively. The expansive evolution proceeds much more slowly in collisions of mass 200 nuclei (in proportion to A$^{1/3}$). Recall that the EOS employed in hydrodynamics stems from lattice QCD \cite{15}, the fundamental theory of the partonic state, which refers to a stationary state.

Concerning the preference for equal mass collision systems we note that the effective center of mass, and thus the position of mid-rapidity, is problematic in "light on heavy" collisions. For the "soft" QCD processes
(such as flow, HBT etc.) it shifts with the impact parameter, toward target rapidity with increasing event centrality. Whereas the effective cm frame, governing the concurrent hard QCD collisions stays fixed at the different midrapidity value of nucleon-nucleon collisions at $\sqrt{s}$. An extreme is presented by the current LHC investigations of p+Pb collisions where soft and hard midrapidity can differ by more that two units \cite{16}, causing interpretational difficulties if one considers the attenuation of hard leading partons with the soft-produced QGP medium and, on top of this, in a narrow rapidity acceptance. Finally, a further practical advantage of A+A collisions is the reflection symmetry about midrapidity. Measurement in one hemisphere suffices, of particular importance in fixed target experiments at the Bevalac, AGS and SPS where the high momentum forward hemisphere was well equipped with state of the art tracking and calorimetric detector techniques.

The physics of hydrodynamic flow did initially require heavy nuclear projectiles. The initially employed, idealized hydrodynamic model \cite{11} illustrated in \Fref{fig1} simply scaled with projectile/target mass A. But the first directed flow results were unambiguously discernible for heavy projectiles only (see \Fref{fig3}).This observation points to the fact that such an idealized model can only resemble the conditions prevailing in a real nucleus-nucleus collision in an ``asymptotic'' manner. In reality only heavy colliding nuclei have a favorable surface to volume ratio and a long-lived fireball. They also featured the high multiplicities required to select centrality classes and to obtain a relatively accurate reconstruction of the reaction plane. Also, the initially employed sphericity method \cite{6} explored only the single particle aspects of the multiparticle distribution whereas later approaches \cite{17,18} exploited the two-and multiparticle correlations
induced by the existence of a reaction plane. Heavy nuclei presented a
problem to the synchrotrons, Bevatron, AGS and SPS: their injector linacs
could only provide for partially stripped "heavy ions", but the modest
synchrotron vacuum (sufficient for proton acceleration) caused further
stripping during acceleration, thus losing the beam. All programs started
with projectiles like $^{20}$Ne or $^{36}$S. The Bevatron required a "vacuum liner"
to produce Au beams in 1985, and at CERN a major vacuum upgrade of the
injector SPS, along with a completely purpose-built source-preaccelerator
complex delivered first Pb beams in 1995.

That much about the early history of directed flow of QCD matter. In the following sections we shall turn to a brief description of the experimental and theoretical development concerning the major different realizations of flow. This development was prompted by the advent of genuinely relativistic projectile energies, from $\sqrt{s}=17$~GeV at top SPS energy via 200~GeV at top RHIC, to the present 2.76~TeV at the LHC. This transition to ``ultrarelativistic'' energies required a fundamental re-formulation of the theoretical framework, while retaining, in principle, the hydrodynamical model for the system dynamical evolution. For the major time span of this evolution a partonic hydro-model is required, employing a QCD plasma equation of state, and amended with initialization models that describe the primordial dynamics occurring during, and right after projectile-target interpenetration, as well as the development of local equilibrium conditions amenable to formulation of an initial energy-momentum tensor, to start the hydro part of the evolution. This was entirely uncharted territory in the early 1990's. Whereas a hadronic hydrodynamics was intuitively plausible (because of the hadronic cross section equalling geometric hadron size), a QCD plasma hydrodynamics was not, and had to be developed, leading to fundamentally new insight as we shall see below. To the more practical side J.Y.Ollitrault demonstrated in a pioneering 1992 publication \cite{17b} that at very high collision energies the longitudinal dynamics decouples from the midrapidity flow generation, thus reducing the relevant coordinate space to the transverse plane: a cylindrical geometry reflecting the onset of boost invariance. Flow initialization was identified with the spatial excentricity $\alpha_{s}$ of the participant density as projected onto the transverse plane. This resulted in the first prediction of in-plane elliptical flow as quantified by the emission anisotropy in transverse momentum space,

\begin{equation} \label{eqn0}
  \bar{\alpha} = \frac{\langle p^{2}(x)\rangle - \langle p^{2}(y)\rangle}{\langle p^{2}(x)\rangle + \langle p^{2}(y)\rangle}
\end{equation}

Ollitrault also noted that if the final anisotropy was to be interpreted as hydro-flow it should characteristically exhibit a proportionality between $\alpha_{s}$ and $\bar{\alpha}$, thus distinguishing it from jet, and other background effects.

The analysis of all different flow patterns in transverse emission space was then unified, and generalized according to a Fourier decomposition of the azimuthal angular emission pattern pioneered by Voloshin and Zhang \cite{18}, and later extended by Poskanzer and Voloshin \cite{5}. It corresponds to ascending moments of the azimuthal emission anisotropy and refers to the triple differential momentum distribution of identified hadrons $i$,

\begin{equation} \label{eqn1}
  \frac{1}{p_{T}} \frac{\dd ^{3}N^i}{\dd p_{T}\dd y\dd \phi}=\frac{1}{2\pi p_{T}} \frac{\dd ^{2}N^i}{\dd p_{T}\dd y} \left\{ 1+2 \sum_{n=1}^{\infty} v_{n}^{i}(p_T,y) \cos[n(\phi^i-\Psi_R)]\right\}
\end{equation}

where $\phi_i$ is the azimuthal emission angle of hadron $i$ and $\Psi_R$ is the angle of the reaction plane. The successive flow moments of order n have the magnitude $v_n$ that depends on $p_T$, rapidity $y$, and impact parameter $b$. The $v_{n}$ are commonly called ``flow coefficients'' thus tacitly implying that their values are, to leading order, a reflection of partonic and (subsequent) hadronic hydrodynamic flow patterns. It took more than a decade of systematic experimental and theoretical study to ascertain this assumption, while also gaining deeper insight about a multitude of non-flow background effects, such as cross-talk among the orders, BE correlations, and low momentum components of jets attenuated in the medium.

The first term, $n=0$, corresponds to an azimuthally symmetric emission pattern, now understood to be caused by the uniform radial flow expansion of the fireball source; it is called radial flow, see \Sref{SectII}. The modern version of directed sideward flow is quantified via determination of $v_1$ (see \Sref{SectIII}). The most exciting recent results concern the elliptic flow $v_2$. Its origins also date back to the Bevalac Plastic Ball experiment and the DIOGENE experiment \cite{18b} where the low energy form
was called "squeeze-out"\footnote{Squeeze-out was later quantified as negative elliptic flow.} \cite{19}, but its full informational content was first appreciated at top RHIC energy, $\sqrt{s}=200$~GeV. It allows for an analysis of one of the most outstanding and important properties of the deconfined QCD plasma, the presence of a certain "minimal" (shear) viscosity \cite{20} that implies a strongly coupled partonic liquid, making contact to recent fundamental field theory, and to results of a "dual" weakly coupled string theory: the so-called AdS$/$CFT interpretation" \cite{21}. \Sref{SectIV} will introduce these ideas.

\begin{figure}[ht]
  \centering
  \includegraphics[width=0.5\textwidth]{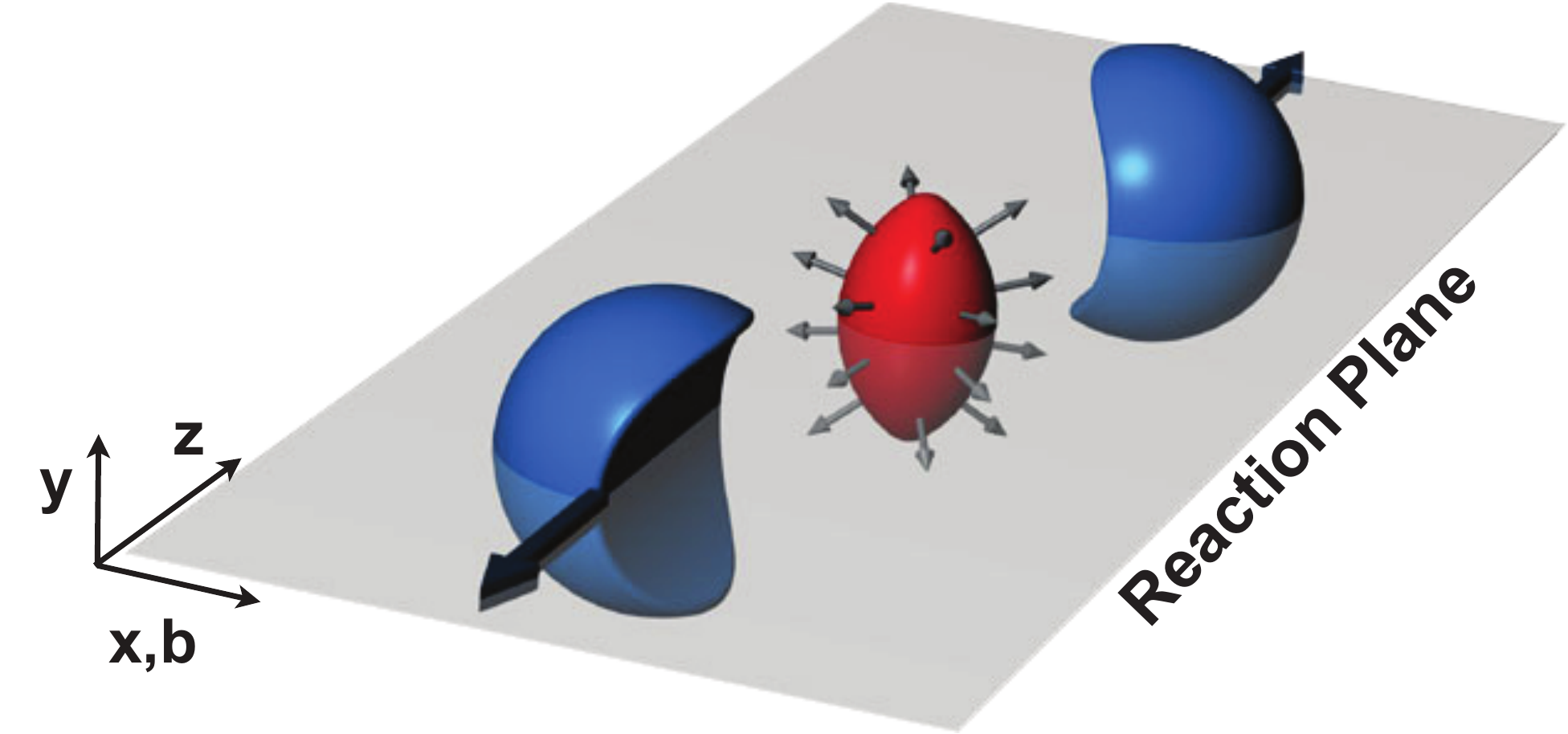}
  \caption{
    Illustration of the origin of elliptic flow in a snapshot of a semi-peripheral collision at top HRIC/LHC energies (ignoring Lorentz contraction). The receding projectile-target residues in the cm frame leave an asymmetric fireball featuring higher gradients of density/pressure along the reaction plane, than out of it.
    \label{fig4}
  }
\end{figure}

Finally, we shall address in \Sref{SectV} a further aspect that is common to $v_{2}$ and all higher flow harmonics. These signals get imprinted into the subsequent flow expansion at primordial time. They are thus sensitive to the very initial dynamics which are addressed by the so-called "color glass condensate" (CGC) theory \cite{23} which investigates a further, novel form of QCD matter governed by gluon saturation. The origin of the elliptic flow signal during primordial conditions is illustrated by the perhaps most often shown plot of recent QGP physics \cite{22} with the schematic illustration of $v_{2}$ initialization at top RHIC and LHC energies, given in \Fref{fig4}. At semi-central impact we see the receding spectator residues of the colliding nuclei, and the central fireball, of elliptical shape. In it the in-plane pressure gradients exceed the out of plane ones: the typical correlation arises here, between position and momentum, that is characteristic of flow fields. This image is drawn ignoring Lorentz contraction but clearly the snapshot must correspond to a "shutter speed" (i.e. time resolution) of about 0.01~fm$/c$, well shorter than the target-projectile interpenetration time which is 0.1~fm$/c$ at top RHIC energy. We see that the $v_{2}$ signal originates early enough to sample properties of the incident state characteristic of CGC formation, and that its dynamical evolution will last long enough to distinguish the presence of a non-viscous ideal hydrodynamic situation from one that would follow due to the existence of a certain amount of shear viscosity.

\begin{figure}[ht]
  \centering
  \includegraphics[width=0.5\textwidth]{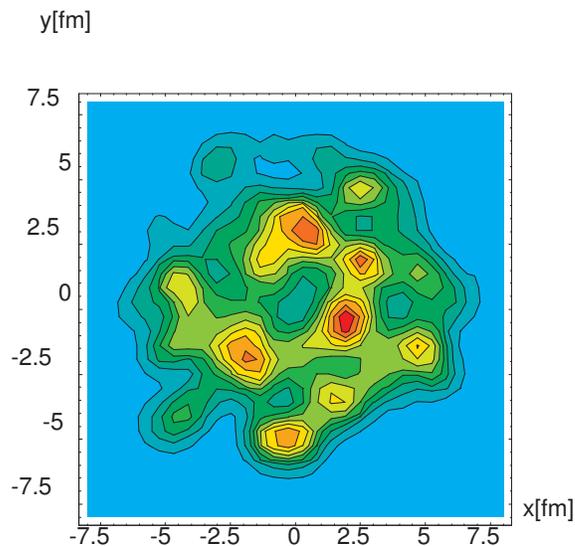}
  \caption{
    Contour diagram of primordial energy density in the transverse plane, created after 0.3~fm$/c$ in a single central Au+Au collision event at $\sqrt{s}=$200 GeV \cite{27}.
    \label{fig5}
  }
\end{figure}

These implications of the $v_{2}$ and $v_{3}$ observables formed a major part of the break-through, accomplished at top RHIC energy: it followed from vastly increased time resolution. Somehow related but still remarkable: an initial gluon density approaching saturation scale leads to extreme multiplicity densities of produced charged hadrons, of order 1500.
This has invited a crucial, further step in event by event analysis: the application of hydrodynamics to individual events \cite{24,25,26}. We shall describe the results in our final \Sref{SectV} but highlight the motivation in \Fref{fig5}. It shows \cite{27} contours of initial energy density projected onto the transverse plane at time 0.3~fm$/c$, calculated from a Monte Carlo model of the target-projectile interpenetration process in a single central Au+Au collision at $\sqrt{s} = 200$~GeV. It indicates a drastic clumpiness prevailing at hydro initialization which is not at all related to impact geometry (which is absent at $b = 0$), in strong contrast to impact geometry playing the major role for the averaged $v_{2}$ initialization as illustrated by \Fref{fig4}. These fluctuations vary from event to event and will introduce fluctuations of the observed $v_{n}$ signals \cite{26,28} which will influence the ensemble averages\footnote{Fluctuations affect the averages because their values are root-mean-square, and therefore increase with fluctuations.} but can also be directly identified by eventwise analysis. Different models of the initial state (such as CGC or Monte-Carlo) will generate different characteristic initial fluctuations. On the other hand, different magnitudes of viscosity will cause different extents of damping such fluctuations \cite{24,26,29}, after the onset of the hydro evolution. The eventwise $v_{2}$ and $v_{3}$ analysis, experimentally and theoretically, might thus finally enable us to disentangle the properties of viscosity, and of the initial state.

\section{Radial Flow} \label{SectII}

\begin{figure}[ht]
  \centering
  \includegraphics[width=0.5\textwidth]{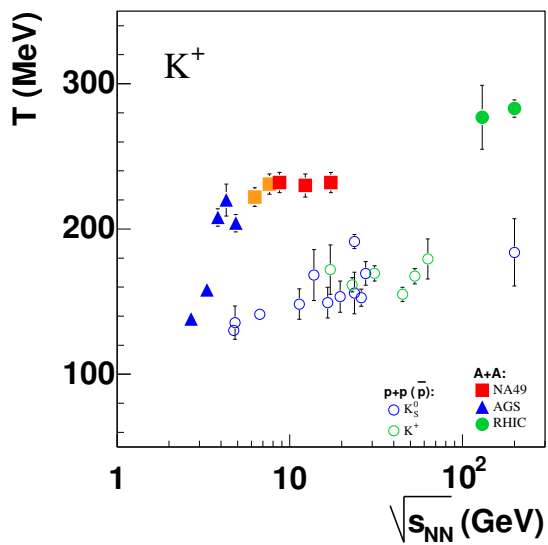}
  \caption{
    The inverse slope parameter T of \Eref{eqn2} for $K^{+}$ transverse mass spectra at p$_{T} <$ 2 GeV$/c$ and midrapidity in central Pb+Pb and Au+Au collisions, and in minimum bias p+p collisions \cite{34}.
    \label{fig6}
  }
\end{figure}

Radial, spherically symmetric expansion flow was first discovered by the FOPI Collaboration at SIS through the linear dependence of the average kinetic energy on the charge of the emitted fragments \cite{29b}. The EOS Collaboration at the BEVALAC \cite{30} observed radial flow in the spectra, following a prediction of "blast wave" expansion by Siemens and Rasmussen \cite{31}. At these low energies this implies an expansion mode of mostly nucleonic matter, setting in after initial compression. In fact Bertsch and Cugnon \cite{32} had shown that such matter expands in a near-isentropic mode, thus confirming an intuitive picture: the nucleon-nucleon total cross section equals the geometrical nucleon size, such that everybody pushes everybody else, into the direction of the pressure and density gradients that point radially in central collisions. In an isentropic expansion the growing spatial volume must be balanced by a concurrent decrease of momentum space volume, but compatible with total energy conservation. Thus, merely lowering the temperature is excluded. The result is a dimensional reduction of momentum space, from a uniform thermal distribution to a 1-dimensional radial ordering of the momenta: collective radial flow. At decoupling (called "kinetic freeze-out") the hadron single particle transverse momentum spectra deviate from the Maxwell-Boltzmann form predicted by the statistical model that envisions an adiabatic expansion of the fireball source \cite{33}:

\begin{equation} \label{eqn2}
  \frac{1}{p_{T}} \frac{\dd^{2} N^i}{\dd p_{T}\dd y} = \frac{1}{m_{T}} \frac{\dd^{2}N^i}{\dd m_{T}\dd y} = a^i \exp( -m^i_{T}/T)
\end{equation}

\noindent where we have changed from the $p_{T}$ to the $m_{T}$ representation,
$m_{T} = \sqrt{p_{T}^{2} + m^{2}}$. We expect this spectral shape to represent the fireball decoupling temperature T in elementary p+p collisions (Hagedorns idea of 1968 \cite{33}), but to fail if one insists to employ it in a central A+A collision if it features an isentropic expansion phase prior to decoupling. The parameter T must thus acquire a different meaning. This expectation is clearly borne out by an analysis of $K^{+}$ inverse slope parameters T (from \Eref{eqn2}) in p+p and Pb+Pb or Au+Au collisions \cite{34}, shown in \Fref{fig6}, for energies ranging from AGS up to RHIC. The p+p inverse slope ascends monotonously up to T of about 160 MeV: the Hagedorn "limiting hadronic temperature" \cite{33}. Whereas the A+A inverse slope increases steeply, to about T = 230 MeV at SPS energy, where a remarkable "plateau" is observed \cite{34}, then onward to T about 280 MeV at RHIC energies. The specific plateau feature, also observed in the corresponding $K^{-}$ spectra \cite{34}, has never been understood conclusively. It has been linked to the hypothesis of a softest point in the EOS at SPS energies, to which we shall turn in \Sref{SectIII}. We note, finally, that the strong increase occurring with top RHIC energies (that continues toward LHC energy) could reflect, in part, a hardening of the $p_T$ spectra by increasing jet production and/or an increasing contribution of pre-hadronic radial flow (see below).

At the decoupling stage the initial hadronization temperature \cite{15}, of about T = 165 MeV, will have fallen down to about 100 MeV. This is the "true" decoupling, "kinetic freeze-out" temperature T$_{F}$. But radial flow has created additional transverse kinetic energy, $m^{i}\langle\beta_T\rangle^{2}$ where $m^{i}$ is the mass of the considered hadron and $\langle\beta_T\rangle$ the average radial flow velocity. We thus expect for the inverse slope parameter of \Eref{eqn2}:

\begin{equation} \label{eqn3}
  T = T_{F} + m^{i} \langle\beta_T\rangle^{2} \mathrm{\qquad at~} p_{T} < 2~\mathrm{GeV}/c,
\end{equation}
and
\begin{equation} \label{eqn4}
  T = T_{F} \sqrt{\frac{ 1 + \langle\beta_T\rangle}{1 - \langle\beta_T\rangle} }  \mathrm{\qquad at~higher~} p_{T}
\end{equation}

\noindent the latter expression indicating a "blue-shifted" \cite{35} decoupling temperature. This arises from the fact that $p_{T}$ and $m_{T}$ are defined relative to the fireball center of mass frame but the actual emission toward the detector occurs from the radial flow frame that is advancing with $\langle\beta_T\rangle$.

\begin{figure}[ht]
  \centering
  \includegraphics[width=0.5\textwidth]{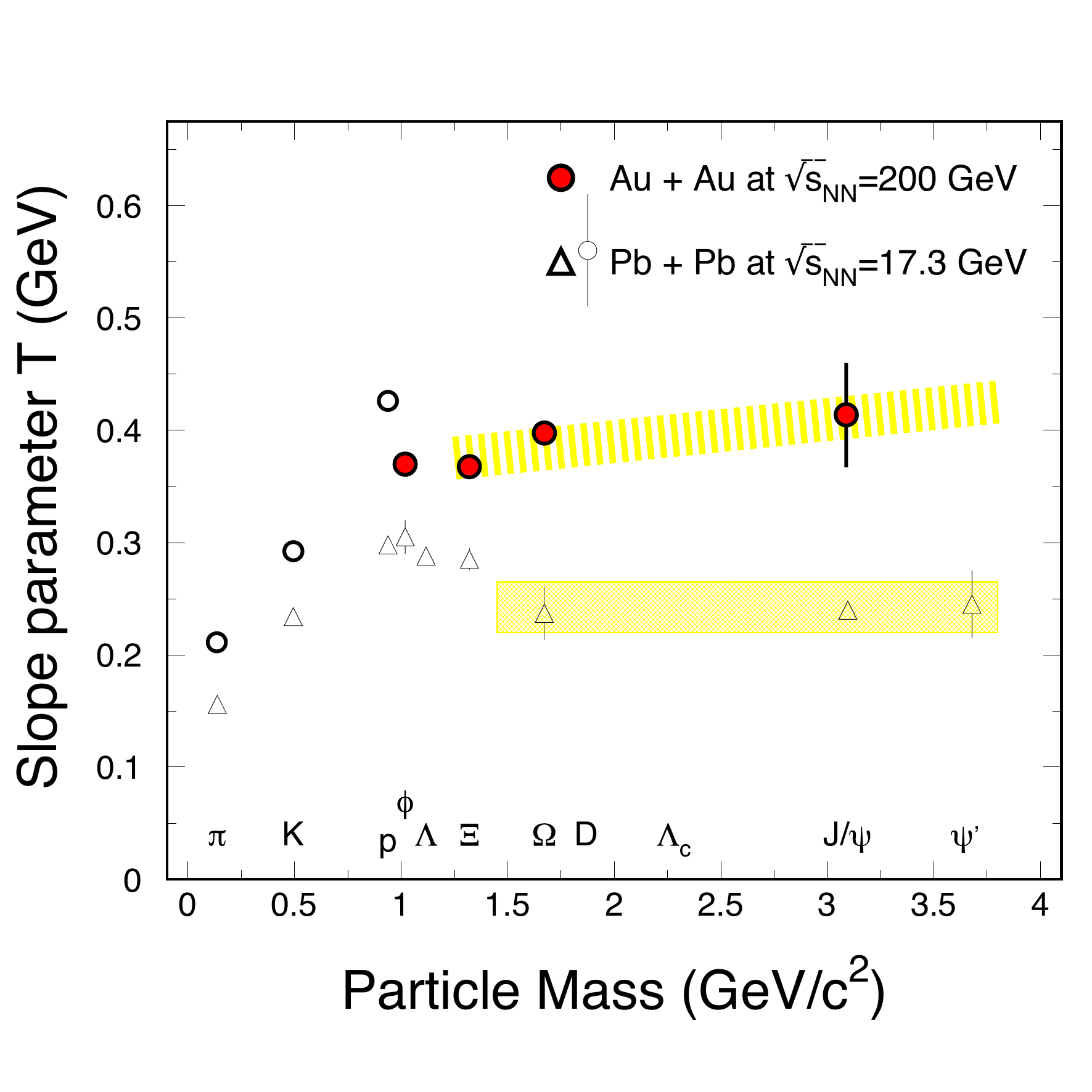}
  \caption{
    Hadron slope parameters T at midrapidity from \Eref{eqn2} as a function of mass. For Pb+Pb at 17.3 GeV, and Au+Au at 200 GeV \cite{36}.
    \label{fig7}
  }
\end{figure}

From \Eref{eqn3} one would expect a simple proportionality of T to the hadron 
mass if radial flow develops, predominantly, during the final hadron/resonance expansion. In reality, a more complex picture is indicated by the data. \Fref{fig7} shows a "Nu Xu" plot \cite{36} - he was the first to realize the implications of such an analysis - giving the systematics of the T parameter for a set of hadronic species produced in central collisions of Pb+Pb at top SPS, and of Au+Au at top RHIC energies.
At both energies one sees a first rise of T, with hadron mass ascending from pions to p, $\Phi$ and $\Lambda$. The RHIC data exhibit a steeper rise, the consequence of a higher radial expansion velocity (see below). However, a drastic change of slope occurs, at both energies, beginning with the hyperons, notably $\Xi$ and $\Omega$, and continuing toward J$/\psi$. At SPS energy T turns down to $\Xi$ and $\Omega$, then stays constant until J$/\psi$. At RHIC the slope diminishes for the hyperons, then slightly ascending to J$/\psi$. Now the latter hadron is forming outside the fireball, and its T slope should thus reflect the constituent charm quark pair flow experienced in the QGP medium: first indications of partonic hydrodynamic flow. 
Concerning the hyperons, in particular the $\Omega$ decouples early from the hadronic expansion due to its small total cross section. Most of its flow should therefore also be of pre-hadronic origin. Taking the above interpretation for granted we finally conclude from \Fref{fig7} that partonic phase flow velocities must be significantly higher at $\sqrt{s}=200$ GeV
than at 17.3 GeV. New LHC data \cite{37} show a further increase at $\sqrt{s}=2.76$ TeV.

The idea of Siemens and Rasmussen \cite{31} to confront single particle transverse momentum spectra of hadrons emitted from an ideal spherically symmetric fireball source that undergoes collective expansion, occurred in 1979: concurrent with the ideas about directed sideward flow. Their "blast wave" equation for the $p_{T}$ spectra first exhibited the well known ambiguity \cite{38} between the freeze-out temperature T$_{F}$ and the surface flow velocity $\beta_T$, not surprising in view of \Eref{eqn3}. We shall show below how to overcome this ambiguity but turn, first, to generalizations of the blast wave description that took account of the fact that, at higher than Bevalac energies, the single spherical fireball description of the source becomes unrealistic due to longitudinal expansion. The rapidity distribution of produced hadrons thus widens toward, at LHC energy, an approximately boost invariant distribution which is flat over a wide domain in rapidity. The first description of radial flow including longitudinal expansion was given by a "hydrodynamically inspired" source parameterization by Schnedermann et al. \cite{35}, resulting in the expression

\begin{equation} \label{eqn5}
  \frac{1}{m_{T}}\frac{\dd N}{\dd m_{T}} = \mathrm{const.} \int_{0}^{R}r \dd r~~m_{T} I_0\left(\frac{p_{T} \sinh(\rho)}{T_F}\right) K_1\left(\frac{m_{T} \cosh(\rho)}{T_F}\right)
\end{equation}

\noindent where R is the transverse surface radius of the source, $\rho = \tanh^{-1} (\beta_T(r))$ and the transverse velocity profile $\langle\beta_T(r)\rangle = 2/3\,\beta_T$(R) $= 2/3\,\beta_T^{max}$.

\begin{figure}[ht]
  \centering
  \includegraphics[width=0.49\textwidth]{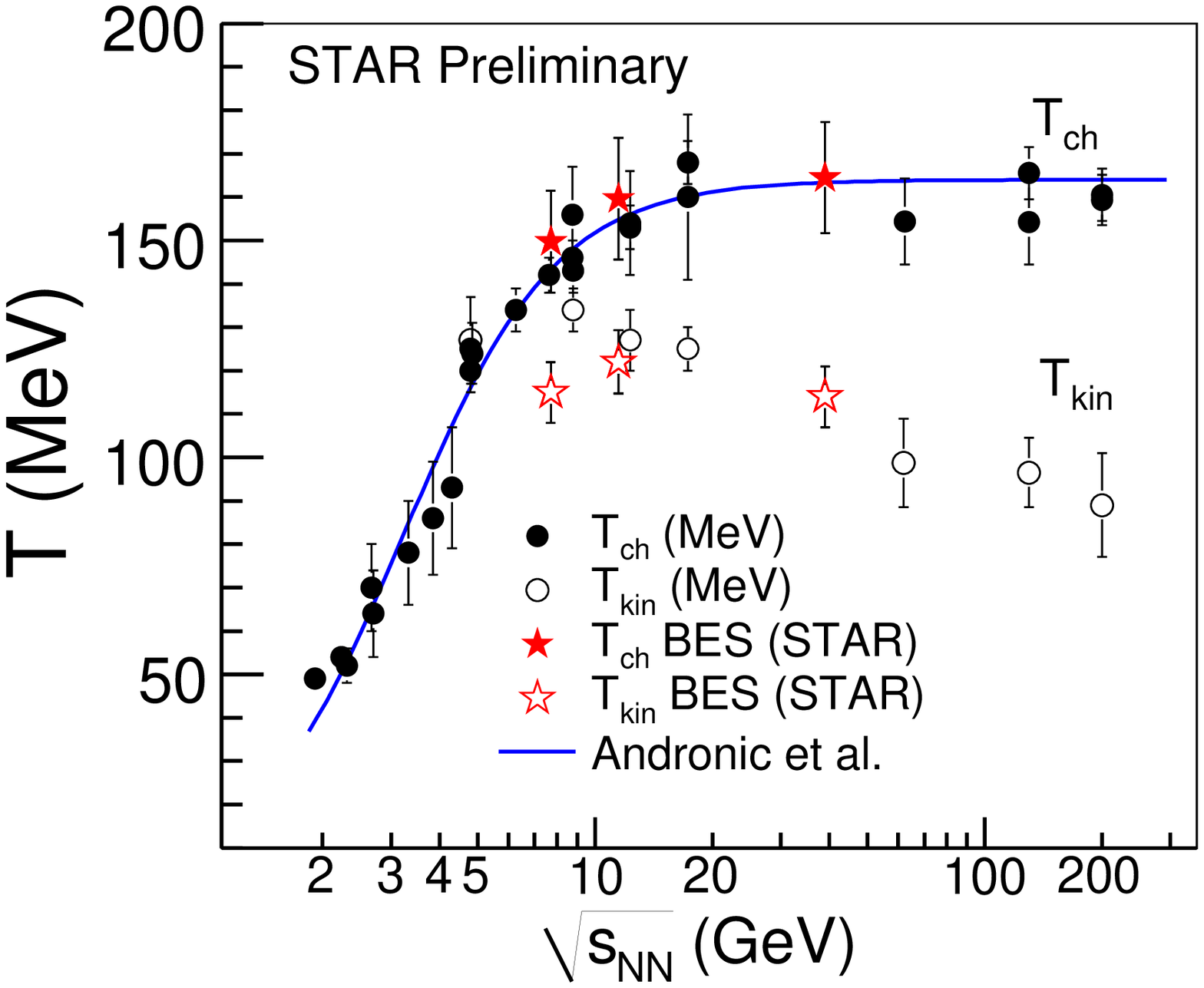}
  \includegraphics[width=0.49\textwidth]{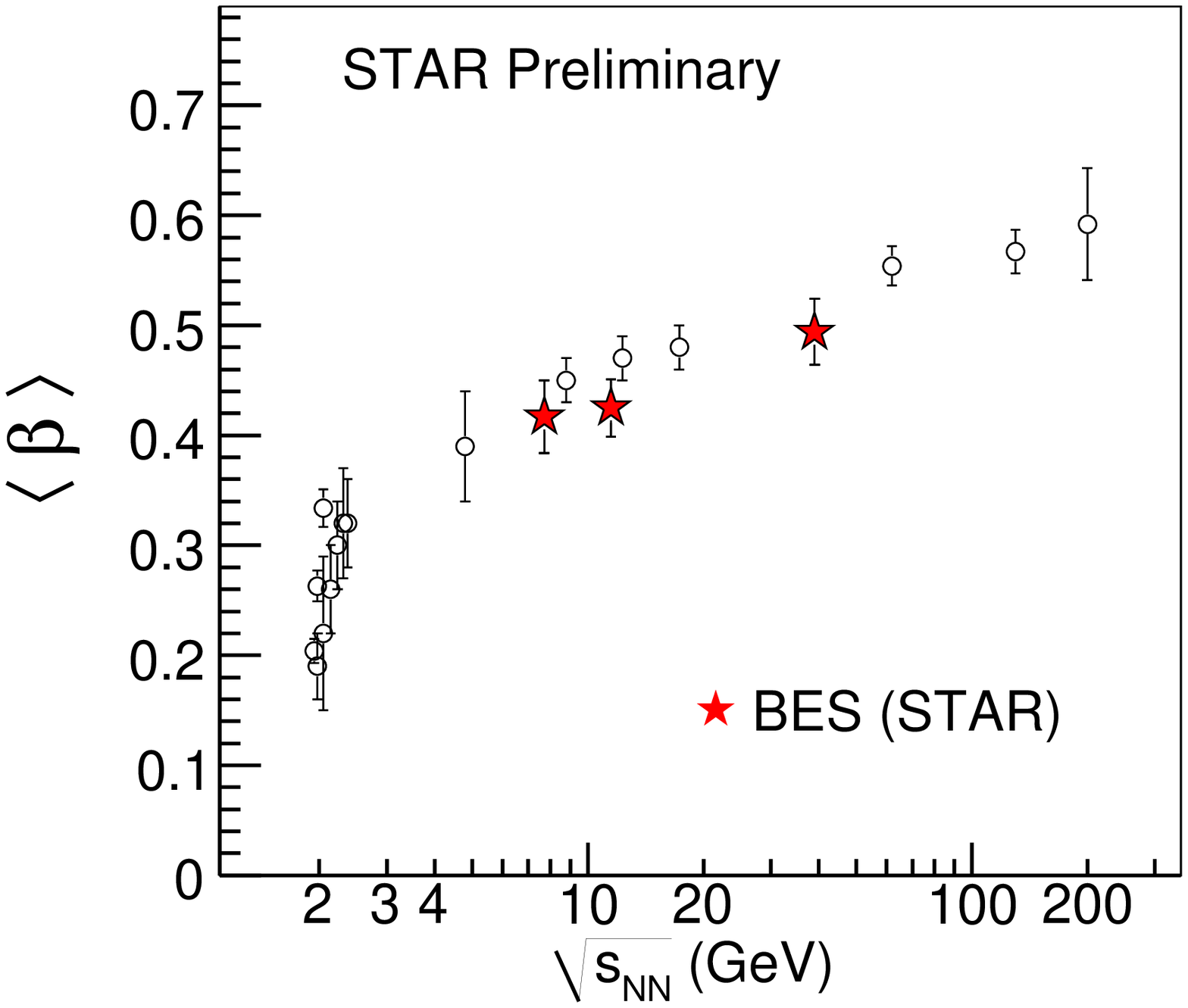}
  \caption{
    (left) The chemical and kinetic freeze-out tmperatures T$_{ch}$ and
    T$_{kin}$ (equal to T$_{F}$ in \Eref{eqn5}), as a function of $\sqrt{s}$. 
    (right) Energy dependence of the average radial flow velocity
    $\langle\beta_T\rangle$ from \Eref{eqn5}. The
    label "STAR preliminary" refers only to the star data symbols \cite{39}. 
    \label{fig8}
  }
\end{figure}

This description of the transverse mass spectra has been widely employed in radial flow analysis, until today. \Fref{fig8} presents a summary \cite{39} of the results up to top RHIC energy. The left hand plot shows the energy dependence of the decoupling (kinetic "freeze-out") temperature T$_{F}$ determined from \Eref{eqn5}, along with the "chemical freeze-out" temperatures T$_{ch}$ that are determined by the Statistical Hadronization Model \cite{40}. It addresses the hadron species multiplicities that are experimentally determined along with the $p_{T}$ or $m_{T}$ spectra. The basic idea of this analysis is, in short, that the observed hadronic multiplicities have become stationary (frozen out) right at the onset of the hadron/resonance expansion stage, due to the smallness of inelastic transmutation among the various species \cite{41,42}. The $p_{T}$ spectra freeze out later, at T$_{F}$, because they reflect the large elastic and resonance cross sections, and the development of radial flow \cite{41}. This feature is clearly born out by the present analysis. While T$_{ch}$ saturates at high energies toward about 165 MeV, the temperature predicted for the parton to hadron confinement transition by Lattice QCD \cite{43}, the kinetic decoupling temperature T$_{F}$ (here called T$_{kin}$) decreases toward about 100 MeV. This latter observation is baffling, at first sight: above $\sqrt{s} = 10$ GeV the chemical freeze-out temperature from which the subsequent hadronic expansion originates, stays practically constant, such that nothing much should change in the subsequent phase. However, we should expect an increasing radial flow velocity imported from the partonic into the hadronic phase. This increases the finally observed surface velocity $\beta_T$. Recalling the anti-correlation in the blast wave formula \Eref{eqn5}, between T$_{F}$ and $\beta_T$ or $\langle\beta_T\rangle$, a flow velocity gradually increased by pre-hadronization flow expansion leads to a gradually smaller apparent T$_{F}$. This interpretation is substantiated by the energy dependence of $\langle\beta_T\rangle$
shown in the right panel of \Fref{fig8}. It grows toward a value of about 0.6. 

\begin{figure}[ht]
  \centering
  \includegraphics[width=0.5\textwidth]{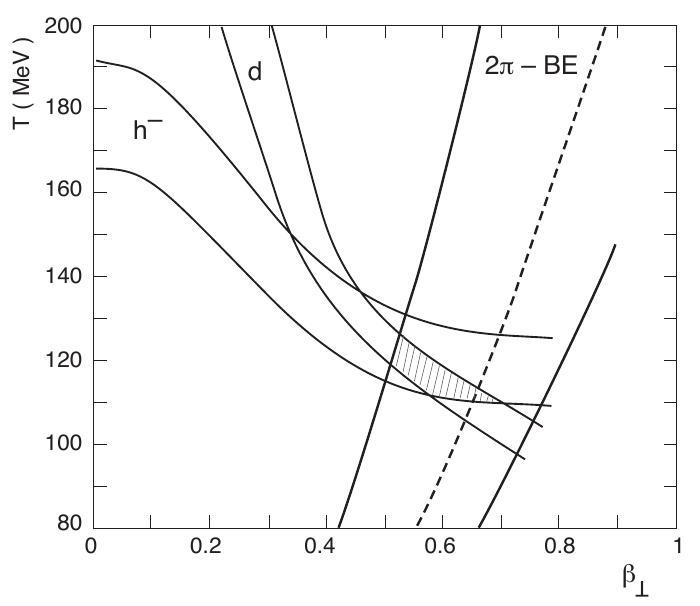}
  \caption{
    Allowed regions of surface freeze-out velocity $\beta_T$ vs. temperature T$_{F}$, in central Pb+Pb collisions at midrapidity and $\sqrt{s}=$17.3 GeV. Combining fits to BE correlations of two negative hadrons, as well as to transverse mass spectra of negative hadrons and deuterons \cite{46}.
    \label{fig9}
  }
\end{figure}

We summarize by observing that radial flow has received relatively little attention recently. Furthermore it is common practice to lift the beta vs. T$_{F}$ ambiguity of \Eref{eqn5} by a simultaneous fit to a wide set of hadronic species. This is not the best one can do. The $m_{T}$ spectra of pions and protons are strongly influenced by resonance decays \cite{45} and, in general, the decoupling occurs not synchronously but sequentially in an expansively diluting medium (in inverse order of the total cross section), such that a hydro or hydro-inspired model may lose its applicability within the course of the hadron/resonance expansion. This calls for the attachment of an "afterburner" phase. Thus
the physics occurring at the parton-hadron phase boundary might be inaccurately reflected in results such as shown in \Fref{fig8}.
A perhaps more reliable determination of the decoupling parameters results from a simultaneous fit of the two-particle Bose-Einstein correlation and the $m_{T}$ spectra of the same particle species \cite{46}. The method is based on Pb+Pb central collision data at 17.3 GeV. Both the BE correlation and the $m_{T}$ spectra have been analyzed employing hydro-inspired source parametrizations \cite{47,48} which give analytic expressions for the BE transverse source expansion radius R$_{out}$
that also exhibit an ambiguity $\beta_T^2/T_{F}$, however under constraints differing from those governing the $m_{T}$ spectral shape. \Fref{fig9} shows 1-sigma error bands which are almost perpendicular to each other, in the T vs. $\beta$ plane (note that we employ here the surface $\beta_T$). This helps lifting the ambiguity. This interesting type of analysis is in some sense superseded now by simultaneous full hydro calculations which, by the way, had to face a decade of "puzzles" \cite{49}, mostly due to a lack of an "afterburner" stage \cite{50}. However, the clearcut effect of a plateau in the inverse $m_{T}$ slopes
seen in \Fref{fig6} stays unexplained.

\section{Directed Sideward Flow}\label{SectIII}

We have already mentioned the discovery of directed hydrodynamic flow of hadronic matter, by the Plastic Ball Collaboration \cite{6}, and illustrated it in \Fref{fig1} to \ref{fig3}. Motivated by the idea \cite{4} to extract the equation of state (EOS) of highly compressed hadronic matter, as the key ingredient of hydrodynamic model application to such data, the study of sideward collective flow has engaged both experiment and theory from 1984 until today. Successive, increasingly refined definitions of this observable resulted in a sequence of approaches toward its quantification: from the "flow angle", illustrated in \Fref{fig2}) to "side flow", and finally to the $v_1$ coefficient in the Fourier decomposition of the eventwise azimuthal emission angles, as described in \Eref{eqn1}. We shall illustrate these developments, from
the Bevalac to RHIC, in detail, but note here that the higher moments of the Fourier decomposition were also getting
initial attention, notably the elliptic flow phenomena quantified by the $v_2$
term in \Eref{eqn1}. At RHIC the elliptic flow signal emerged
as a key signal sensitive to the
primordial (hydro-)dynamics of partonic matter (see \Sref{SectIV}).

\begin{figure}[ht]
  \centering
  \includegraphics[width=0.5\textwidth]{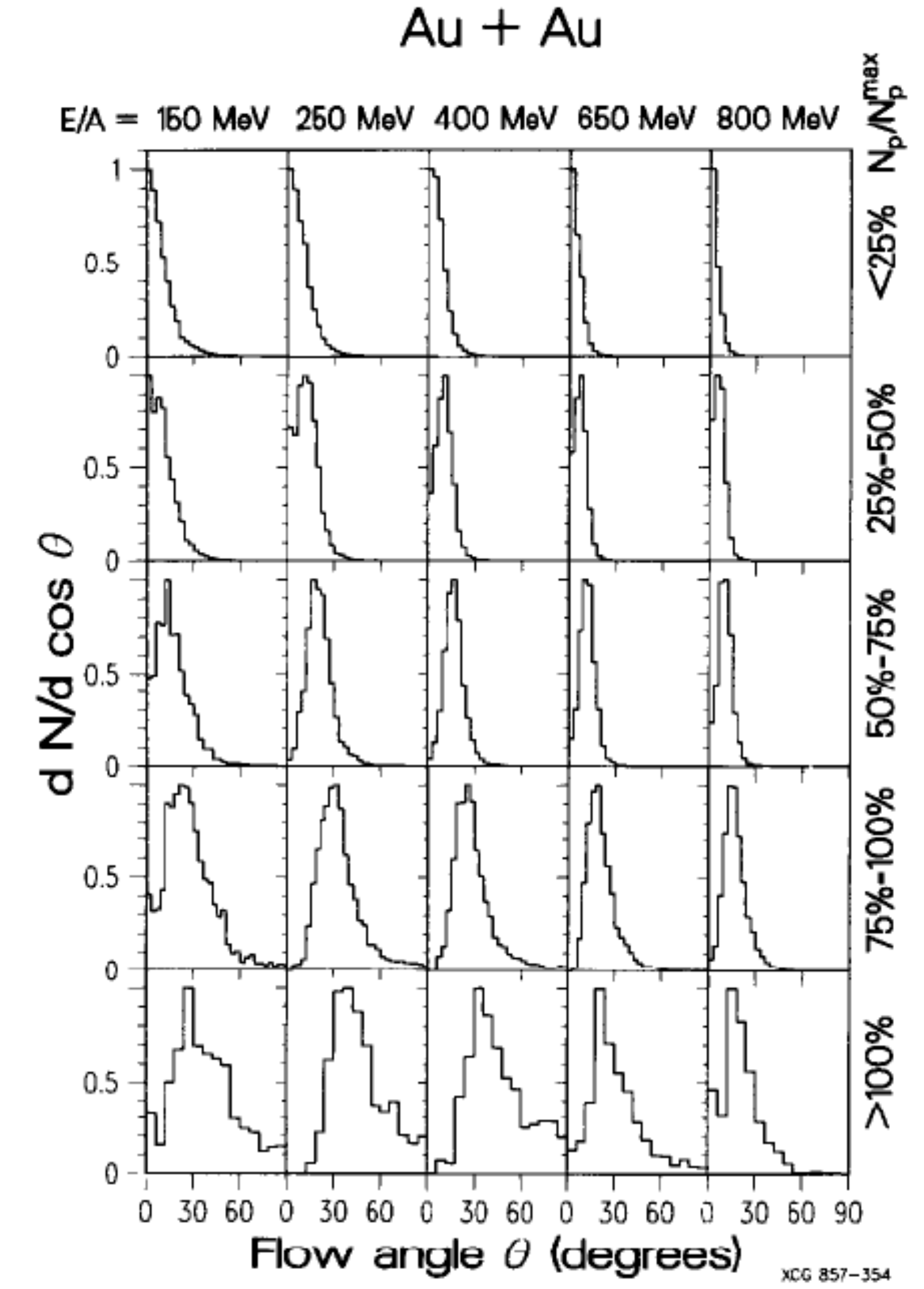}
  \caption{
    The flow angle from sphericity analysis at the Bevalac \cite{7}, for Au+Au collisions at incident energies from 150 to 800 MeV per nucleon, in proton multiplicity windows from peripheral to central events.
    \label{fig10}
  }
\end{figure}

Returning to the discovery of directed flow at the Bevalac, we show in \Fref{fig10} the results of the Plastic Ball Collaboration study \cite{7} of the flow angle (\Fref{fig2}) in Au+Au collisions at energies ranging from 150 to 800 MeV per projectile nucleon ($A$ MeV), and for successive multiplicity (centrality) cuts. The flow angle histograms result from the so-called sphericity analysis \cite{51}. Here, the event shape in momentum space gets determined by the flux tensor,

\begin{equation} \label{eqn6}
  F(i,j) = \sum_{n=1}^{M} p_i(n)p_j(n)/2m(n)  \qquad \mathrm{for~i,j = x,y,z}
\end{equation}

\noindent where n denotes the particle, with mass m and cartesian momentum components $p_i$ and $p_j$, and the sum goes over all M particles in the given event. The eigenvalues, and eigenvectors define the global event shape. Note that the factor $1/2m(n)$ represents a weight; as chosen here \cite{51} F(i,j) represents the kinetic energy distribution in the event. The kinetic energy flow ellipsoid (see \Fref{fig1} and \ref{fig2}) points away from the beam axis with the flow angle Theta: this is the discovery! Its principal axis defines the reaction plane, together with the beam direction.

This intuitive "hydrodynamically inspired" interpretation sees the sideward flow as a consequence of the finite impact vector that characterizes the initial collisional geometry: similar to a collision of two water droplets. The energy density in the projectile-target overlap zone results from the stopping-down of the initial longitudinal beam kinetic energy, to internal degrees of freedom (compression and thermal excitations) of the fireball. The pressure gradient is in-plane, pushing fireball and spectator matter sideward. The degrees, both of initial stopping-down of the participant matter, and of the resulting sideward pressure gradients, are the non-trivial ingredients of this, otherwise geometry dominated picture. One clearly perceives the first two elements of a theoretical description of the collisional evolution: the initialization period setting the stage for the hydro-flow, and the EOS that governs it. Via these two factors the initial geometrical constellation gets translated into a corresponding flow pattern. This also applies to the higher moments of the thus imprinted primordial fireball energy density distribution, as illustrated in \Fref{fig4} for the origin of elliptic flow, and implied in the Fourier decomposition of \Eref{eqn1} for even higher moments of the finally observed flow pattern.

As our introductory review precedes an entire volume of more focused treatises concerning the state of the art in flow analysis, we shall treat ourselves here to two general considerations:

1. Implicit in the above, intuitive proposal, of initial state geometrical constellations resulting in finally detected modes of particle emission from flow fields, there is a strictly nontrivial assumption: that the dynamical evolution can, in fact, be described by a hydro model that incorporates the microscopic conditions (viscosity, EOS etc.) of a hadronic or partonic medium in small size A+A collisions. And that this hydro evolution preserves the imprint at initialization, mapping it to the observables of the final state, perhaps with some amount of blurring that would reveal the presence of initial fluctuations and/or viscosity. In the present example of directed sideward flow the geometry plays such an overarching role that it appears almost inevitable (\Fref{fig1}) to recover it in the final state. However, going to higher moments the geometrical pre-conditions become more subtle, and their respective fate during evolution might provide for tests of increasing sensitivity. It is clear a priori that the initial state fireball source can be decomposed into Fourier moments. Likewise it is clear that the final emission pattern can be similarly analyzed. The key task is to arrive at a mapping of the source development, moment by moment. Such a study has been pioneered by the RHIC PHOBOS Collaboration \cite{52}, see \Sref{SectV}.

2. With the increasing resolution of successive modes in a Fourier decomposition of the fireball source, it becomes clear that the dominating role of the impact geometry diminishes. The strong primordial energy density fluctuations in the source, as illustrated in \Fref{fig5}, will start to dominate the signal. They will also become imprinted into the eventwise flow field, as an initial condition in addition to impact geometry, or even instead of it. Note that an ideal zero impact parameter event, like the one shown in \Fref{fig5}, has no non-isotropic aspects of geometrical origin
but should still exhibit strong non-isotropic fluctuations, the magnitude and pattern depending on the physics that governs the initialization period, such as the decay of a primordial color glass condensate (CGC) state to a subsequent QGP flow evolution. Will these structures also be mapped to the final state? The answer will depend on the magnitude of viscosity, among others. Selection of "super-central" events, and the so-called "event shape engineering" \cite{53} will answer this question (affirmatively \cite{54}) at RHIC and LHC, as we shall show in \Sref{SectV}.

\begin{figure}[ht]
  \centering
  \includegraphics[width=0.45\textwidth]{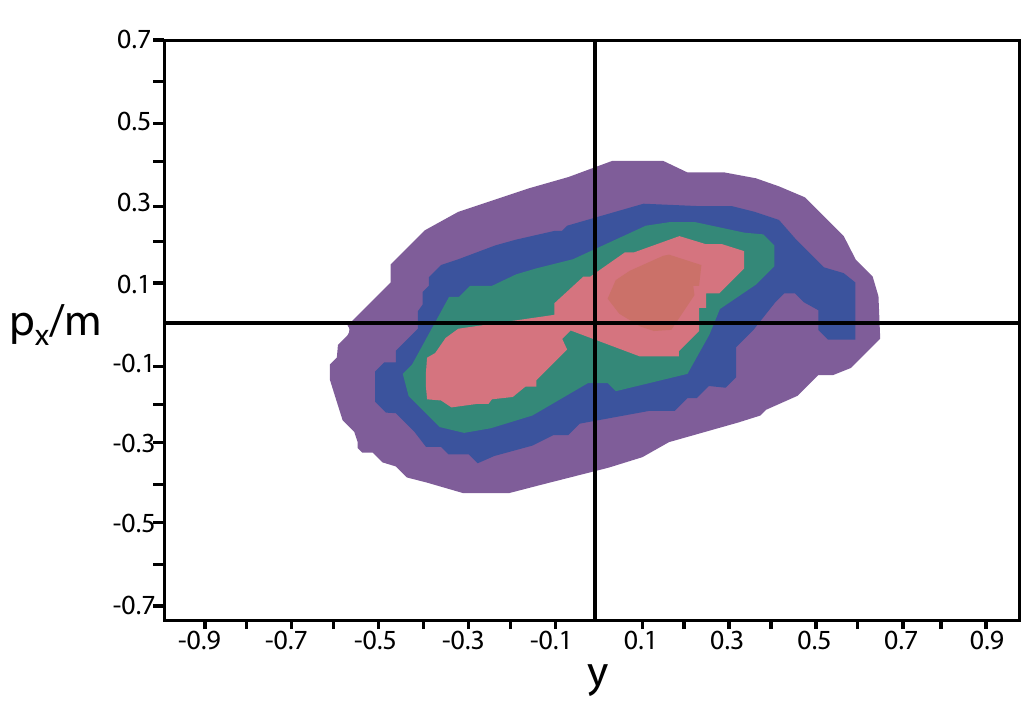}
  \includegraphics[width=0.65\textwidth]{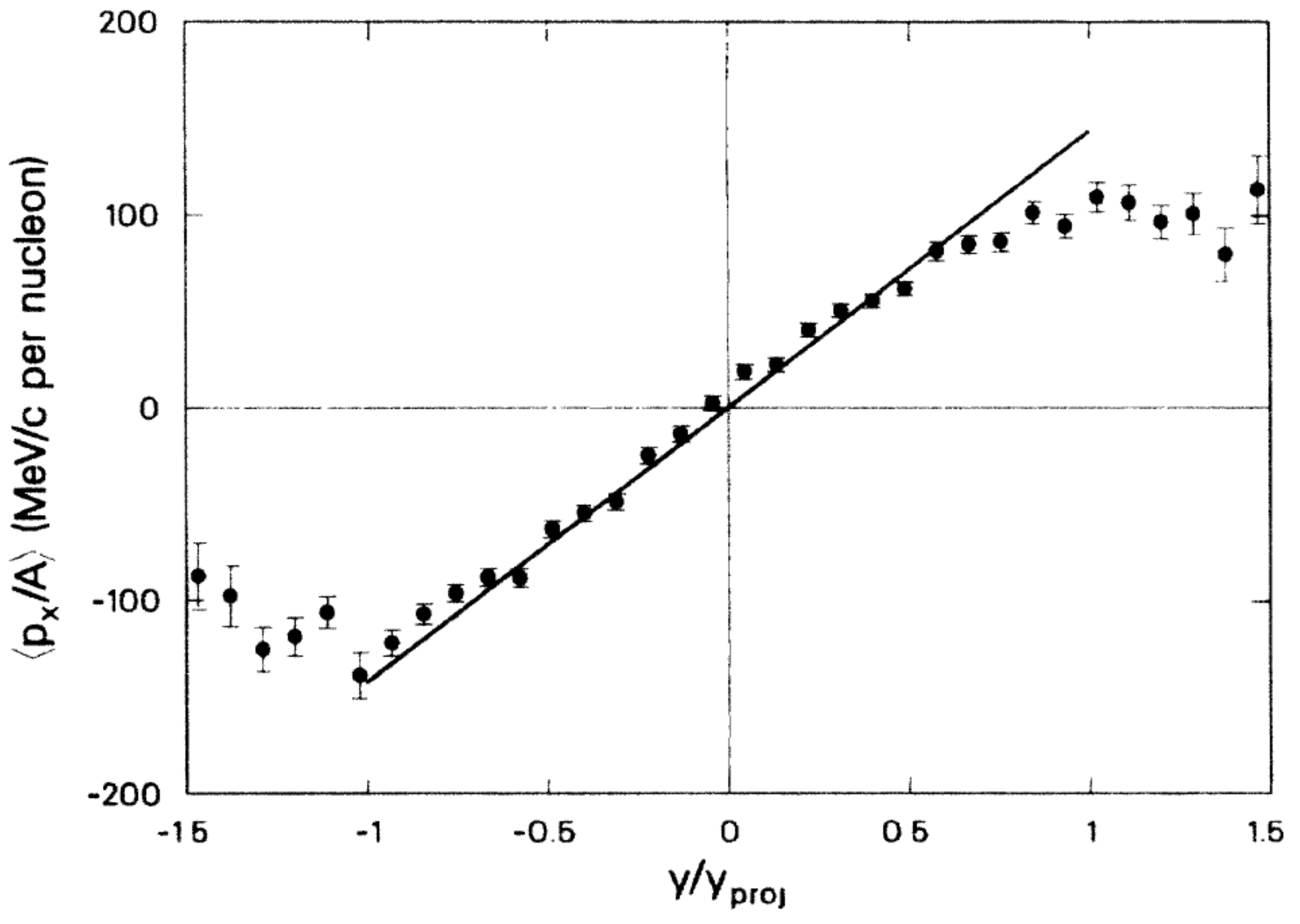}
  \caption{
    Illustration of the event plane analysis \cite{17} in a Plastic Ball study of directed flow \cite{8}. Top panel: Contour diagram of transverse proton momenta projected onto the eventwise reaction plane, $p_x/m$, vs. proton rapidity, in semi-central Nb+Nb collisions at 400$A$ MeV. Bottom panel: the average $\langle p_x \rangle$ per nucleon in successive slices of cm proton rapidity as scaled by the projectile rapidity.
    \label{fig11}
  }
\end{figure}

Returning to the further development of sideward flow we note, first, that the sphericity analysis turned out to be highly sensitive to detector features like tracking efficiency/accuracy, and acceptance. This encouraged the development of alternative sideflow analysis methods. The next step, the so-called "event plane analysis" developed by Danielewicz and Odyniec \cite{17}, is illustrated in \Fref{fig11} by a Plastic Ball investigation \cite{8} of Bevalac Ca+Ca, Nb+Nb and Au+Au collisions at various energies and centralities.

For each event the reaction plane angle $\Psi_{R}$ gets determined, as a first step, by constructing the event "direction vector"

\begin{equation} \label{eqn7}
  \vec{Q} = \sum_{n} \omega  p_T(n) \times  ( \cos(\phi), \sin(\phi) )
\end{equation}

\noindent where the weight $\omega$ is +1 in the forward rapidity hemisphere, and -1
at backward rapidity, and $p_T(n)$ denotes the transverse momentum of particle n which has azimuthal angle $\phi$. The direction of $\vec{Q}$, and the beam direction z, define the reaction plane angle

\begin{equation} \label{eqn8}
  \Psi_{R} = \arctan (Q_y/Q_x).
\end{equation}

\noindent More accurately, the reaction plane is estimated by $\vec{Q}$, within the experimental resolution.

As the next step, all transverse momenta of protons in the event are projected onto its reaction plane, yielding entries $p_x$/m (where m is the proton mass) at the particle rapidity y. Finally, all entries from all events of the analyzed ensemble can be gathered in one plane, $p_x/m$ vs. rapidity $y$. The top panel of \Fref{fig11} shows the resulting contour plot, for the reaction Nb+Nb at 400 MeV per nucleon. Taking the average $p_x$ in slices of rapidity, $\langle p_x\rangle$ or $\langle p_x\rangle/m$, one arrives at the characteristic plot shown in the bottom part of \Fref{fig11}. 

\begin{figure}[ht]
  \centering
  \includegraphics[width=0.5\textwidth]{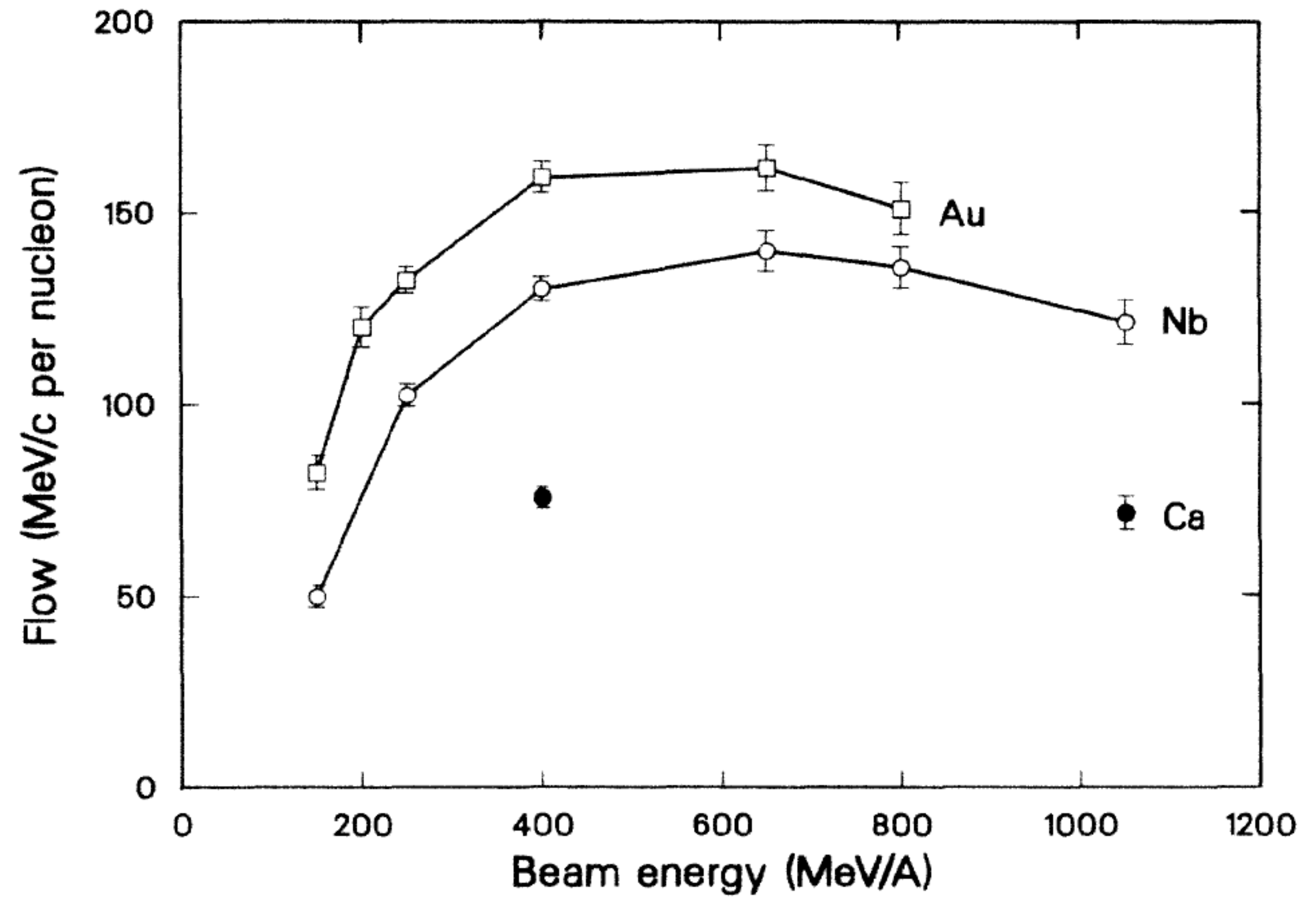}
  \caption{
    Summary of Plastic Ball data for Ca+Ca, Nb+Nb and Au+Au collisions \cite{8}. The flow observable $F(y)$ from \Eref{eqn9} is shown vs. the beam energy.
    \label{fig12}
  }
\end{figure}

The shape of this plot demonstrates the collective transfer of transverse momentum between the target- and projectile nucleons. The slope quantifies the intensity of this momentum flow:

\begin{equation} \label{eqn9}
  F(y) = d\langle p_x/A\rangle/dy
\end{equation}

\noindent is now called sideflow and has dimension MeV$/c$ per nucleon. Evidently $Q$ is only an estimate for the true reaction plane, and thus the projections $p_x$ into this plane are too small. A correction factor can be obtained \cite{17} by splitting each event into two equal size subevents. The correlation of the subevent plane angles $\Psi_A$ and $\Psi_B$ gives the resolution:

\begin{equation} \label{eqn10}
  res = \sqrt{2} \times \sqrt{\langle cos(\Psi_A-\Psi_B)\rangle}
\end{equation}

\noindent and the final result thus is $p_x$ = $p_x^{observed}$/res. This formalism leads to the data summary of this Plastic Ball analysis \cite{8} that we show in \Fref{fig12}. The flow from \Eref{eqn9} increases with the projectile mass. Note that this method can also quantify the flow for the light Ca+Ca collision system, for which the sphericity analysis \cite{6} shown in \Fref{fig3} had yielded, at most, a marginal signal.

\begin{figure}[ht]
  \centering
  \includegraphics[width=0.5\textwidth]{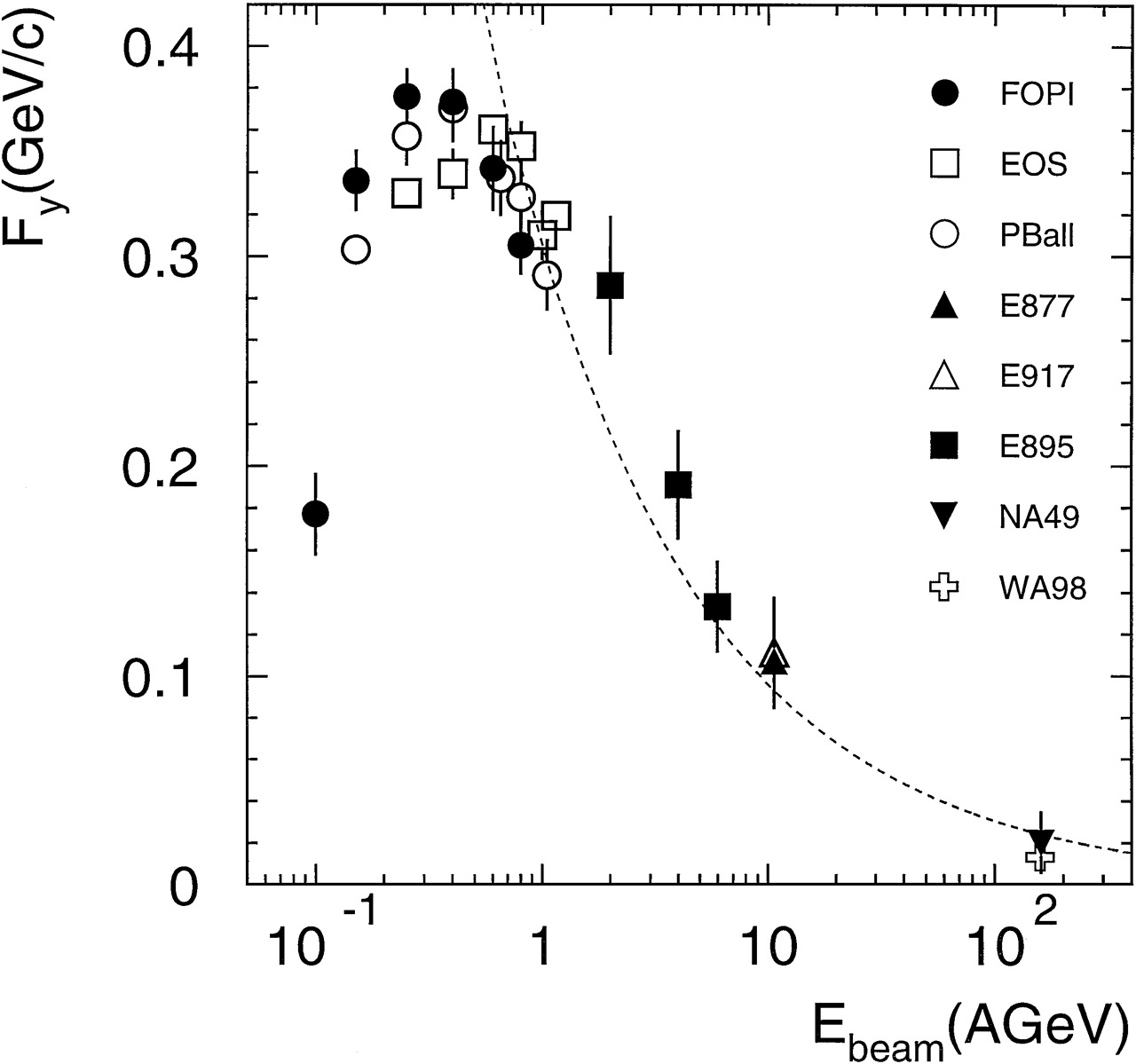}
  \caption{
    Excitation function of proton sideflow $F(y)$, from Bevalac to SPS energies \cite{55}. The dashed line is based on \Eref{eqn11} assuming a constant force acting on the target-projectile interface area during passage time.
    \label{fig13}
  }
\end{figure}

More than a decade later these data had been extended by many other experiments (EOS, FOPI, E877, E917, E895, NA49 and WA98), at the Bevalac, at SIS18 of GSI, at the AGS(BNL) and SPS(CERN), the latter offering Pb beams of 158 GeV/nucleon ($\sqrt{s}$ = 17.3 GeV). \Fref{fig13} gives a 1999 synopsis \cite{55} of the results for F(y). All data included here refer to intermediate impact parameters, b = (0.4-0.6)$b_{max}$, where the directed flow phenomena are most clearly expressed as we saw from \Fref{fig1}, already.

In all cases the slope parameter F(y) was obtained by a fit of the curve in the $p_x$ vs. rapidity plot over a wide range in y as seen in the bottom part of \Fref{fig11}. Potential small scale effects in the direct vicinity of mid-rapidity, to which we shall turn below, are ignored here.

\begin{figure}[ht]
  \centering
  \includegraphics[width=0.5\textwidth]{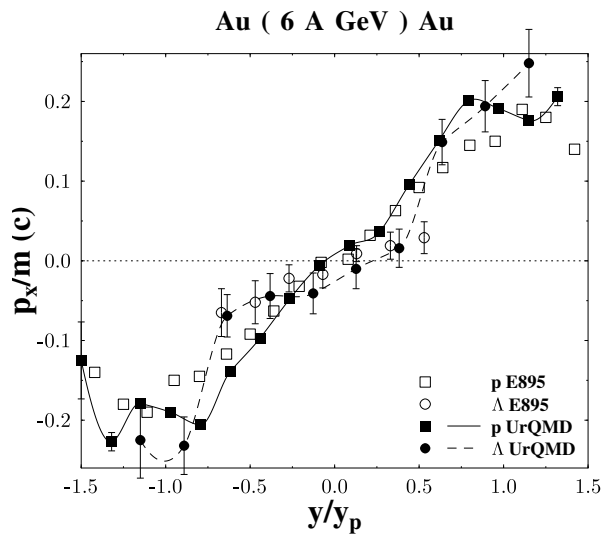}
  \caption{
    The average $\langle p_x \rangle/m$ from event plane analysis of semi-central Au+Au collisions at 6$A$ GeV \cite{56}, for proton and Lambda particles, compared to UrQMD predictions for $b < 7$ fm \cite{57}.
    \label{fig14}
  }
\end{figure}

The main impression from \Fref{fig13} is a steep decrease of directed flow toward high energies. A qualitative explanation \cite{55,56} follows from the consideration
that the momentum transferred collectively to the participating nucleons
results from a force field developing in the fireball that acts by deflecting
nucleons sideward, during the passage time $t_{pass}$ of the target and
projectile density distributions. In fact, the quantity F(y)/$t_{pass}$ has
the dimension of a force, such we can write (somewhat symbolically)

\begin{equation} \label{eqn11}
  F(y) = P_{eff} \times S \times t_{pass}
\end{equation}

\noindent where $S$ denotes the area of the effective interface, and $P_{eff}$ the average pressure, exerted over the interface, during the passage time.
Qualitatively, we thus gain access to the pressure which depends on the fireball density and on the EOS. The passage time decreases linearly with $\sqrt{s}$ but the energy density increases much more slowly with $\sqrt{s}$.
The dashed line shown in \Fref{fig13} is based on the radical assumption of a constant force $P_{eff} \times S$ acting during the passage time. Above Bevalac/SIS18 energies the overall trend of the data is well described by a decreasing passage time (see ref. \cite{55} for detail). Such an effect should clearly also be borne out in a microscopic transport model description of directed flow data. As an example we show in \Fref{fig14} directed flow $\langle p_x\rangle/m$ data for protons and Lambda particles from an E895 AGS study of semi-central Au+Au collisions at 6$A$ GeV \cite{57}, compared to a UrQMD calculation of the Frankfurt group \cite{56} which gives a good overall description.

\begin{figure}[ht]
  \centering
  \includegraphics[width=0.49\textwidth]{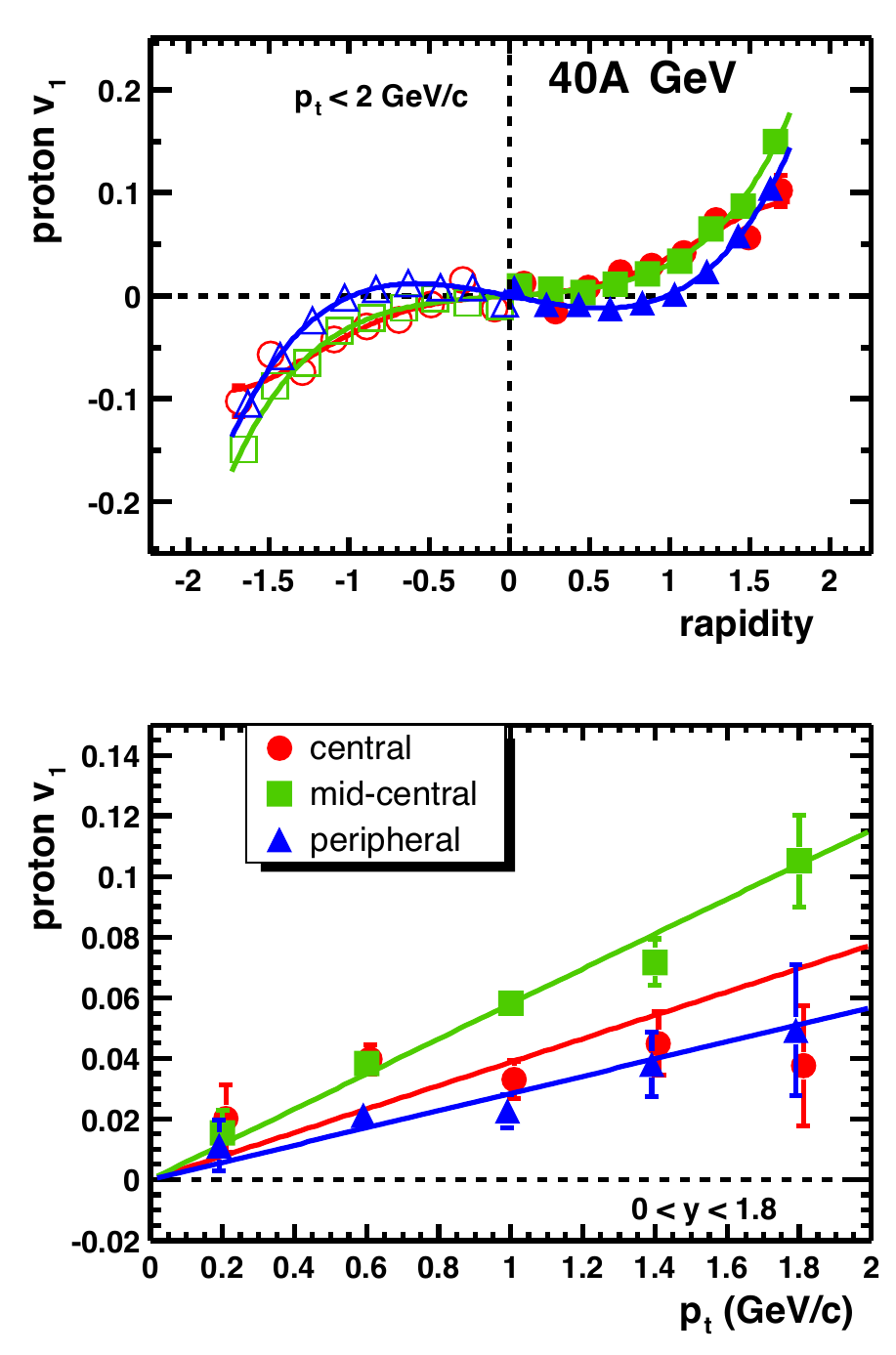}
  \includegraphics[width=0.49\textwidth]{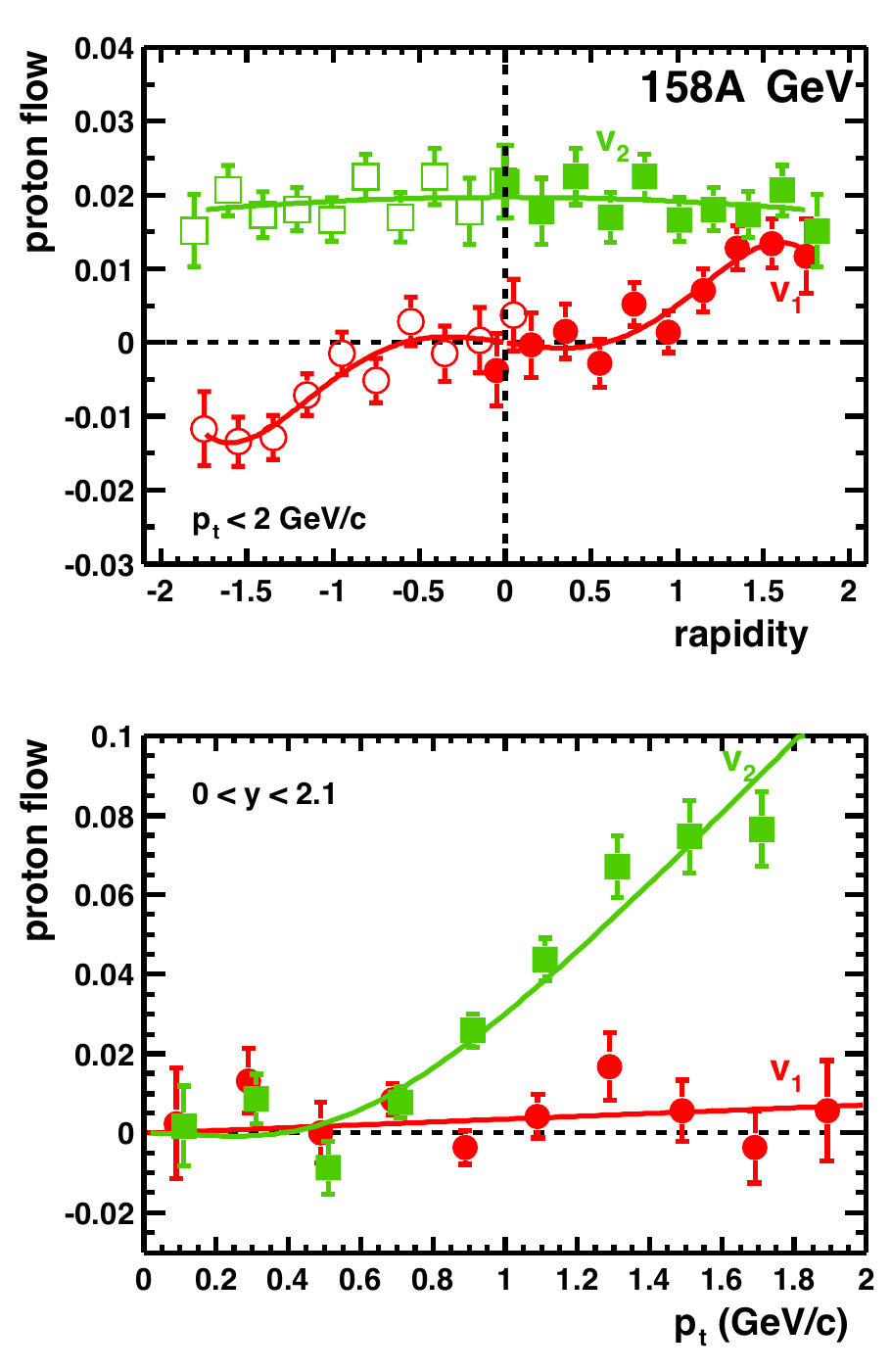}
  \caption{
    The flow coefficient $v_1$ for protons in semi-central Pb+Pb collisions at the SPS energies 40 and 158$A$ GeV \cite{58}, as a function of cm rapidity, and of transverse momentum.
    \label{fig15}
  }
\end{figure}

\begin{figure}[ht]
  \centering
  \includegraphics[width=0.5\textwidth]{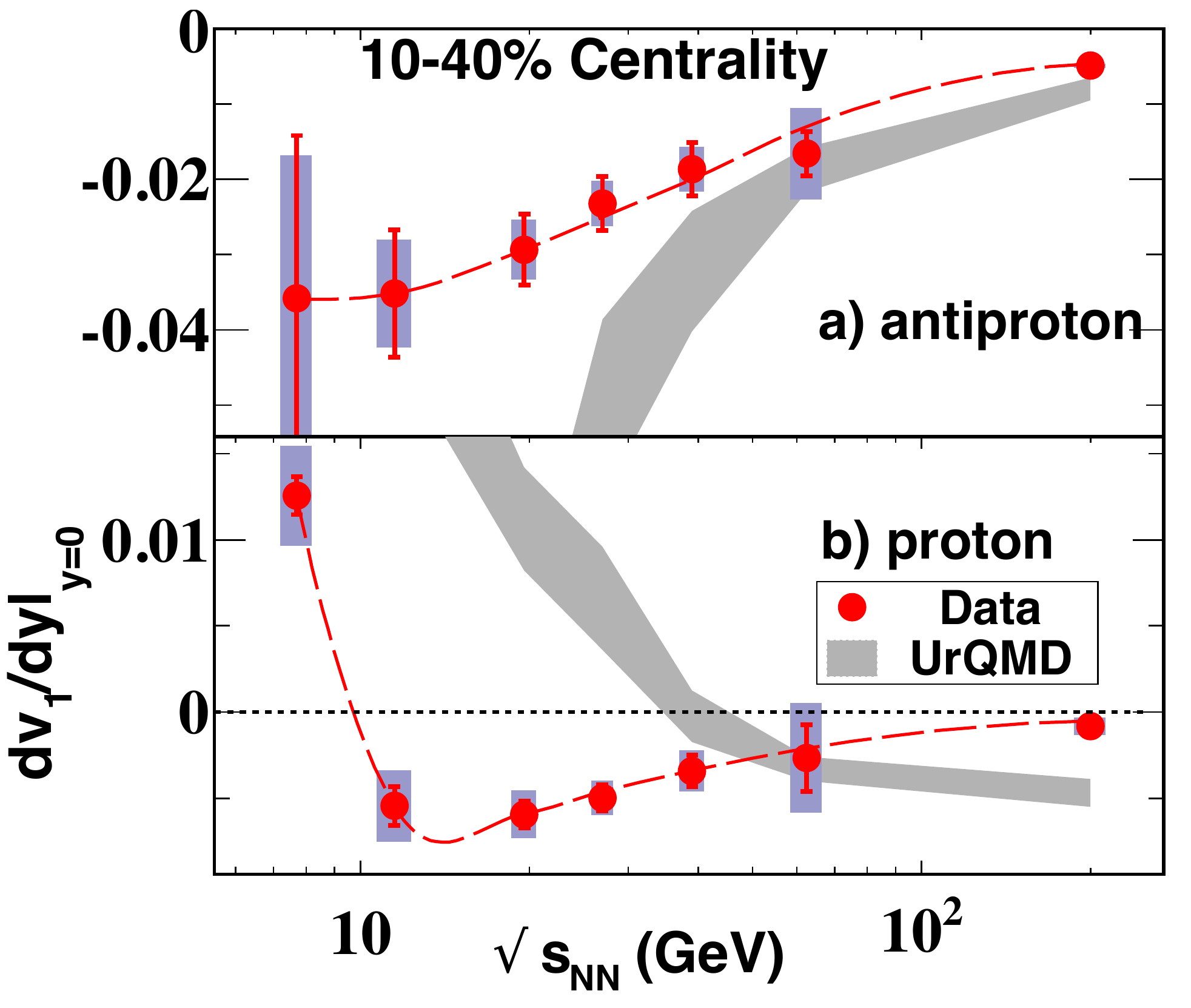}
  \caption{
    Directed proton and antiproton flow slope d$v_1/$d$y$ at midrapidity as a function of beam energy, for intermediate centrality Au+Au collisions at RHIC energies from 7.7 to 200 GeV \cite{60}. The bands show the corresponding UrQMD predictions. The dashed curves are drawn to guide the eye.
    \label{fig16}
  }
\end{figure}

Later on, however, the NA49 Collaboration \cite{58} arrived at a detailed study of the rapidity dependence near midrapidity. In Pb+Pb collisions at two SPS energies, 40 and 158$A$ GeV ($\sqrt{s}$ = 8.73 and 17.3 GeV), they observed a significant change of the y dependence, away from a simple s shape, to a domain with inverted slope: the so-called "collapse of the flow signal" \cite{59}. \Fref{fig15} shows the proton $v_1$ coefficient with negative slope d$v_1$/dy. This observation has recently been confirmed, in more detail, by the RHIC STAR collaboration \cite{60}. We show in \Fref{fig16} their results for the midrapidity slope of $v_1$ in semi-central Au+Au collisions over a wide range of energies, $\sqrt{s}$ = 7.7 to 200 GeV. Proton midrapidity "antiflow" is observed from 10 GeV to 200 GeV, where the signal approaches zero, along with the $v_1$ coefficient, itself.

It should be noted that the flow description has changed here, from the flow measure F(y) of \Eref{eqn9} to the $v_1$ coefficient in the Fourier decomposition of \Eref{eqn1}. Whereas the former is a quantity in transverse momentum space, the latter is based entirely on the distribution of azimuthal emission angle in the cm frame, without reference to the magnitude of the corresponding momenta. It is, thus, dimensionless,

\begin{equation} \label{eqn12}
  v_1 = \langle p_x\rangle/\langle p_T\rangle
\end{equation}

\noindent and corresponds to the mean value $\langle cos(\phi - \Psi_R)\rangle$ from reaction plane analysis \cite{5,55}.

With Figures \ref{fig13} and \ref{fig16} we have covered proton directed flow from Bevalac to RHIC energies. Also shown in \Fref{fig16} is the STAR result \cite{60} for antiprotons
which exhibit "antiflow" throughout. We mention in passing that this is the case also for pions and kaons \cite{55}. This is plausible intuitively: with an ellipsoidal baryonic source pointing into the flow direction the emissivity for newly created particles points oppositely, in the reaction plane.

In a hydrodynamic description of flow (radial, sideward and elliptic) the EOS defines the pressure that we implied in \Eref{eqn11}. Pressure vanishes in an ideal first order phase transition which occurs at a "critical" temperature $T_c$ and density $\epsilon_c$. In a real A+A collision we can not describe the dynamical evolution by a universal, single trajectory in ($T$, $\epsilon$) but by a band of such trajectories because of the fireball inhomogeneities. Thus an A+A collision can not make a point landing on
($T_c$,$\epsilon_c$) which would then occur at a sharply defined incident
energy, but would be influenced by the phase transition over a certain energy domain within which the average EOS pressure stays low: the dynamical band would contain the "softest point". Moreover, one would expect \cite{61} that the effect is maximal when the incident energy is just sufficient for the dynamical evolution to enter the partonic QCD phase (the "onset" energy domain \cite{34}). At much higher energy the evolution begins at $T > T_c$ and the system is already in full expansion when crossing the critical values, thus "racing" through the transition. This idea has been most persistently pursued by the Frankfurt theory group \cite{59,62}. 

We see from \Fref{fig16} that a transport model like UrQMD (shaded bands) which has no phase transition does most distinctly fail to describe the proton antiflow result. The UrQMD d$v_1$/d$y$ also turns negative above about 50 GeV but this should be a trivial consequence of the fact that at such high energies most of the protons stem from newly created proton-antiproton pairs which produce antiflow as the top panel of \Fref{fig16} shows. However this can not explain the data at 10 GeV where the $\overline{\mathrm{p}}$/p ratio is about 0.05 only. Surprisingly these observations have not found a quantitative theoretical explanation as of yet. Regarding hydrodynamics, the exciting developments with the higher flow modes, at top RHIC and LHC energies, have attracted most if not all of the attention. And the hydro model early time initialization methods have all been tuned to the corresponding situation that target-projectile interpenetration (or $t_{pass}$) is ultra-fast (short) and occurs well before the startup of a hydro evolution. At low energies the situation is fully complementary, interpenetration takes about 3 fm$/c$ at $\sqrt{s}$ = 10 GeV but the hydro phase must begin well earlier. The system is at no time well synchronized, some parts being in flow already whereas others have not even engaged in the collision. This might call for a return of the former 3-fluid models which were so happily abandoned.

\section{Elliptic Flow}\label{SectIV}
\begin{figure}[h!]   
\begin{center}
\includegraphics[scale=0.6]{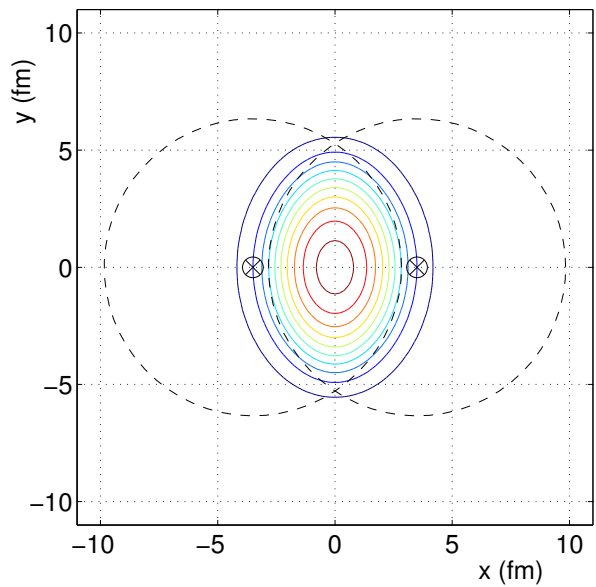}\vspace{-0.3cm}
\caption{Transverse projection of primordial binary collision density in an
Au+Au collision at impact parameter $7 \:fm$, exhibiting participant
parton spatial eccentricity \cite{63}.}
\label{fig17}
\end{center}
\end{figure}

\begin{figure}[h!]   
\begin{center}
\includegraphics[scale=0.55]{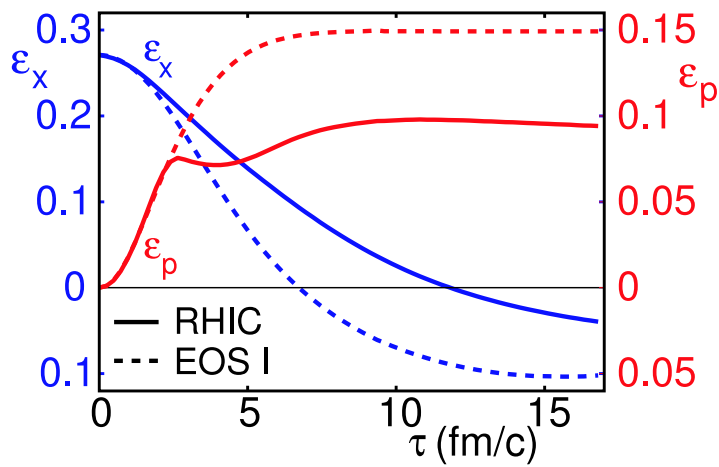}\vspace{-0.3cm}
\caption{Time evolution of the spatial eccentricity $\epsilon_x$ and the
momentum space anisotropy $\epsilon_p$ (equations~\ref{eqn13} and \ref{eqn14}) in
the hydrodynamic model of an Au+Au collision at $b=7 \: fm$,
occurring at $\sqrt{s}=200 \: GeV$ \cite{63}. The dynamics is
illustrated with two equations of state.}
\label{fig18}
\end{center}
\end{figure}

Before turning to the details of elliptic flow data we wish to
illustrate \cite{63} the hydrodynamic model view of \Fref{fig4}. \Fref{fig17} exhibits the
transverse projection of average primordial energy density, assumed to be
proportional to the number density of participant nucleon collisions in the
overlap volume arising from a Au+Au collision at impact parameter
$b=7 \: fm$. The nuclear density profiles (assumed to be of
Woods-Saxon type) intersect in an ellipsoidal fireball, with minor
axis along the direction of $\vec{b}$ which is positioned at $y=0$.
The obvious geometrical deformation can be quantified by the spatial
eccentricity (unfortunately also labeled $\epsilon$, like the energy density, in the
literature)
\begin{equation}
\epsilon_x(b)=\frac{\left<y^2-x^2\right>}{\left<y^2+x^2\right>}
\label{eqn13}
\end{equation}
where the averages are taken with respect to the transverse density
profiles of \Fref{fig17}. $\epsilon_x$ is zero for $b=0$, reaching a value
of about 0.3 in the case $b=7 \: fm$ illustrated in \Fref{fig17}.\\
 \\
Translated into the initialization of the hydrodynamic expansion
the density anisotropy implies a corresponding pressure
anisotropy. The pressure is higher in $x$ than in $y$ direction, and
thus is the initial acceleration, leading to an increasing momentum
anisotropy,
\begin{equation}
\epsilon_p(\tau)=\frac{\int dxdy \: (T^{xx}-T^{yy})}{\int dxdy \:
(T^{xx}+T^{yy})}
\label{eqn14}
\end{equation}
where $T(x,y)$ is the fluid's energy-momentum tensor.
\Fref{fig18} shows \cite{63,64} the time evolution of the spatial and
momentum anisotropies for the collision considered in \Fref{fig17},
implementing two different equations of state which are modeled with
(without) implication of a first order phase transition in ''RHIC''
(''EOS1''). A steep initial rise is observed for $\epsilon_p$, in
both cases: momentum anisotropy builds up during the early partonic
phase at RHIC, while the spatial deformation disappears. I.e. the
initial source geometry, which is washed out later on, imprints a
flow anisotropy which is preserved, and observable as ''elliptic
flow''. A first order phase transition essentially stalls the
buildup of $\epsilon_p$ at about $\tau=3 \: fm/c$ when the system
enters the mixed phase, such that the emerging signal is almost
entirely due to the partonic phase. We have to note here that the
''ideal fluid'' (zero viscosity) hydrodynamics \cite{64} employed
in \Fref{fig18} is, at first, a mere hypothesis, in view also of the fact
that microscopic transport models have predicted a significant
viscosity, both for a perturbative QCD parton gas \cite{65,66} and
for a hadron gas \cite{67,68}. Proven to be correct by the data, the ideal fluid description of 
elliptic flow tells us that the QGP is a non-perturbative liquid
\cite{69,70}.\\
\begin{figure}[h!]   
\begin{center}\vspace{-0.2cm}
\includegraphics[width=0.99\textwidth]{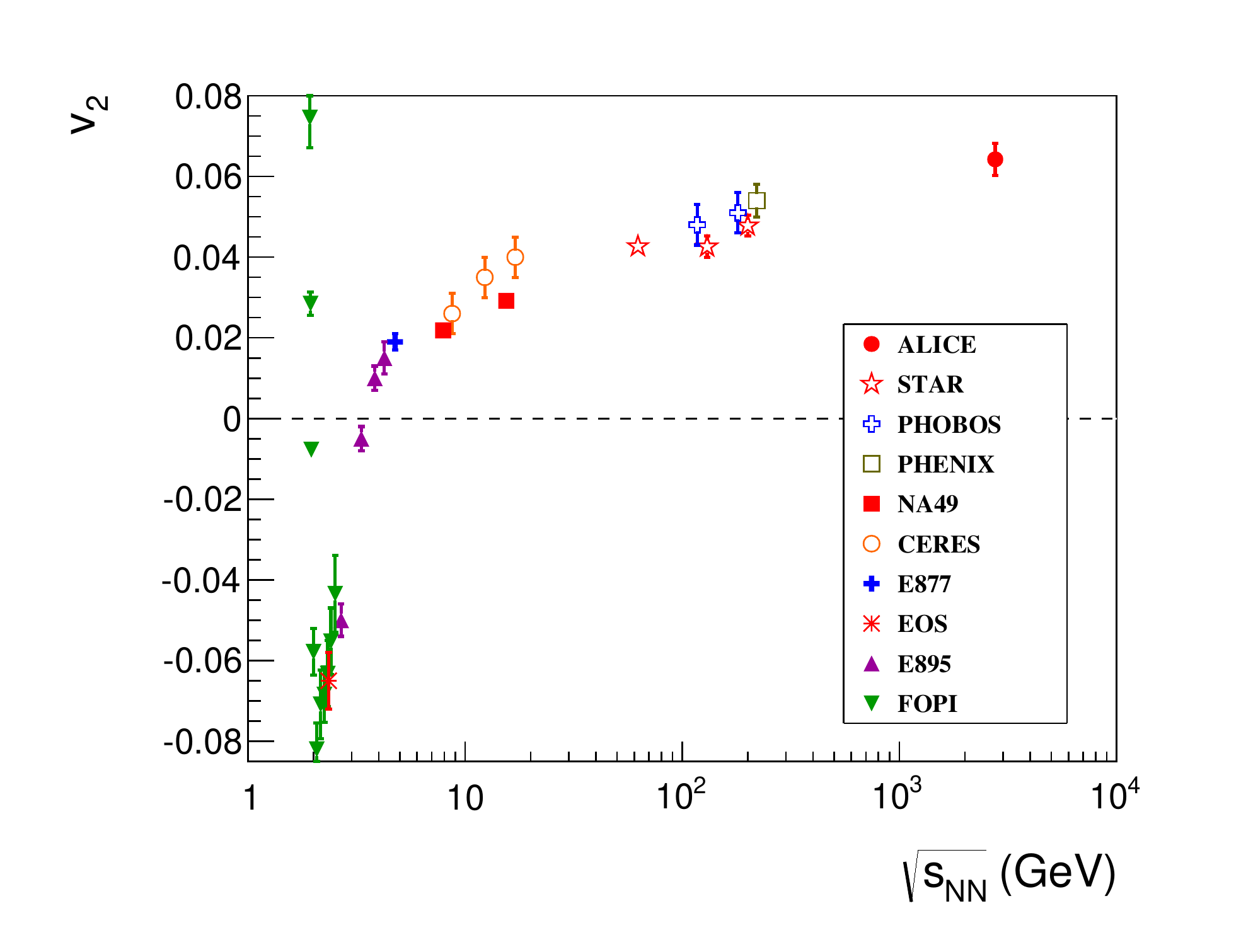}\vspace{-0.4cm}
\caption{Energy dependence of the elliptic flow parameter $v_2$ at
mid-rapidity and averaged over $p_T$, in Au+Au and Pb+Pb
semi-peripheral collisions \cite{71}.}
\label{fig19}
\end{center}
\end{figure} \\
Elliptic flow is quantified by the coefficient $v_2$ of the second
harmonic term in the Fourier expansion (see \Eref{eqn1}) of the invariant cross
section; it depends on $\sqrt{s}, b, y$ and $p_T$. \Fref{fig19} shows the
$\sqrt{s}$ dependence of $v_2$ at mid-rapidity and averaged over
$p_T$, in Au+Au/Pb+Pb semi-peripheral collisions \cite{58,71,72,73,73a,73b}. We 
see that the momentum space anisotropy exhibits a rich structure. As a function of cm energy it changes from in-plane emission to out-of-plane emission, "squeeze-out", and back to in-plane emission. We note that the pattern of $v_2$ energy dependence in \Fref{fig19} reflects substantial influences that are not related to hydrodynamic flow. In particular, the dramatic excursion to negative $v_2$ can be ascribed to a trivial kinematical effect of semi-peripheral collision geometry: the ``cold'' spectator subvolumes of the colliding system are not removed in longitudinal direction, unlike in the ultra-relativistic situation that is addressed in \fref{fig4}. They block the emission from the participant volume, in the in-plane direction. A mere kinematical effect that disappears, gradually, over the $\sqrt{s}$ interval from about 3 to 6~GeV, due to increasing Lorentz contraction both of the source, and of the spectator remnants. Thus ``anti-flow'' should not be related to hydrodynamic flow, about which we can not learn anything due to this occulting of the source. Unlike the final increase of the $v_2$ signal, onward from SPS energy, to which we shall turn below, in the discussion of the ``hydrodynamic limit'' associated with \Fref{fig22}.\\

\begin{figure}[h!]   
\begin{center}\vspace{-0.5cm}
\includegraphics[scale=0.45]{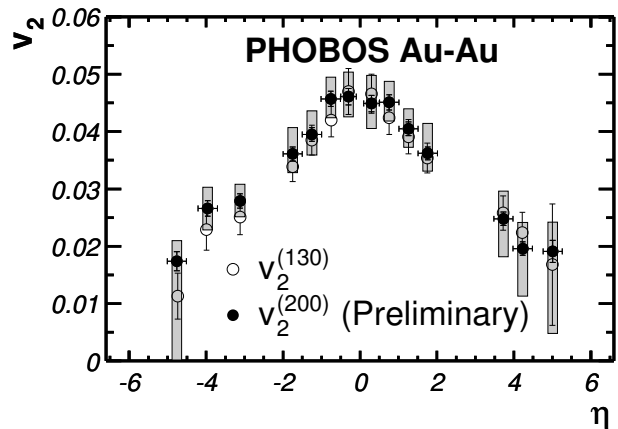}\vspace{-0.4cm}
\caption{Pseudo-rapidity dependence of the $p_T$-averaged elliptic flow
coefficient $v_2$ for charged hadrons at $\sqrt{s}=130$ and $200 \:
GeV$ \cite{74}.}
\label{fig20}
\end{center}\vspace{-0.2cm}
\end{figure}
\Fref{fig20} shows the (pseudo)-rapidity dependence of $v_2$ at $\sqrt{s}=130$ and 200
$GeV$ as obtained by PHOBOS \cite{74} for charged particles in
minimum bias Au+Au collisions. It resembles the corresponding
charged particle rapidity density distribution, suggesting
that prominent elliptic flow arises only at the highest attainable
primordial energy density, in the vicinity of mid-rapidity. \\
\begin{figure}[h!]   
\begin{center}\vspace{-0.5cm}
\includegraphics[scale=0.45]{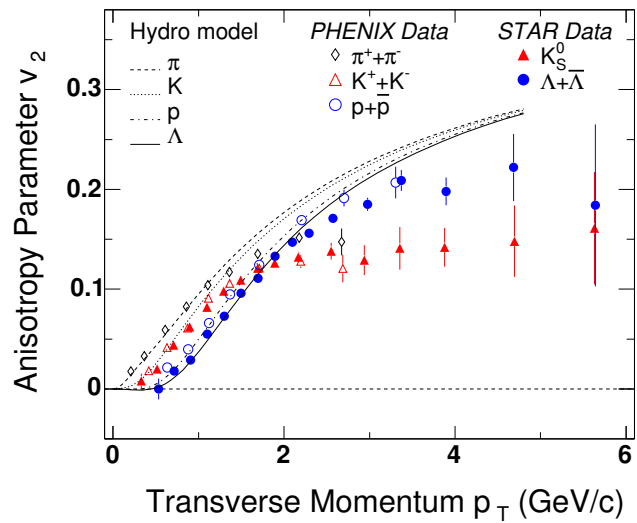}\vspace{-0.4cm}
\caption{Transverse momentum dependence of elliptic flow $v_2$ for mesons and
baryons in Au+Au collisions at $\sqrt{s}=200 \: GeV$. The
hydrodynamic model \cite{63} describes the mass dependence at
$p_T \le 2 \: GeV/c$.}
\label{fig21}
\end{center}
\end{figure} \\
That such conditions are reached at top RHIC energy is shown in \Fref{fig21} and~\ref{fig22}. The former combines STAR \cite{75} and PHENIX \cite{76} data for the 
$p_T$ dependence of elliptic flow, observed for various identified hadron species
$\pi, K, p, K^0$ and $\Lambda, \overline{\Lambda}$ in Au+Au at
200 $GeV$. The predicted hydrodynamic flow pattern \cite{63,77}
agrees well with observations in the bulk $p_T  < 2 \: GeV/c$
domain. \Fref{fig22} (from \cite{58}) unifies average $v_2$ data from AGS
to top RHIC energies in a scaled representation \cite{72} where
$v_2$ divided by the initial spatial eccentricity $\epsilon_x$ is
plotted versus charged particle mid-rapidity density per unit
transverse area $S$, the latter giving the density weighted
transverse surface area of primordial overlap, \Fref{fig17}. \Fref{fig22}
includes the hydrodynamic predictions \cite{67,63,64,77,78} for
various primordial participant or energy densities as implied by the
quantity $(1/S)dn_{ch}/dy$ \cite{72}. Scaling $v_2$ by $\epsilon_x$
enhances the elliptic flow effect of near-central collisions where
$\epsilon_x$ is small, and we see that only such collisions at top
RHIC energy reach the hydrodynamical flow limit. The latter is illustrated here for models that include an EOS ansatz which incorporates \cite{63} the effect of a
first order phase transition, which reduces the primordial flow
signal as was shown in \Fref{fig18}. \\
\begin{figure}[h!]   
\begin{center}
\includegraphics[width=0.5\textwidth]{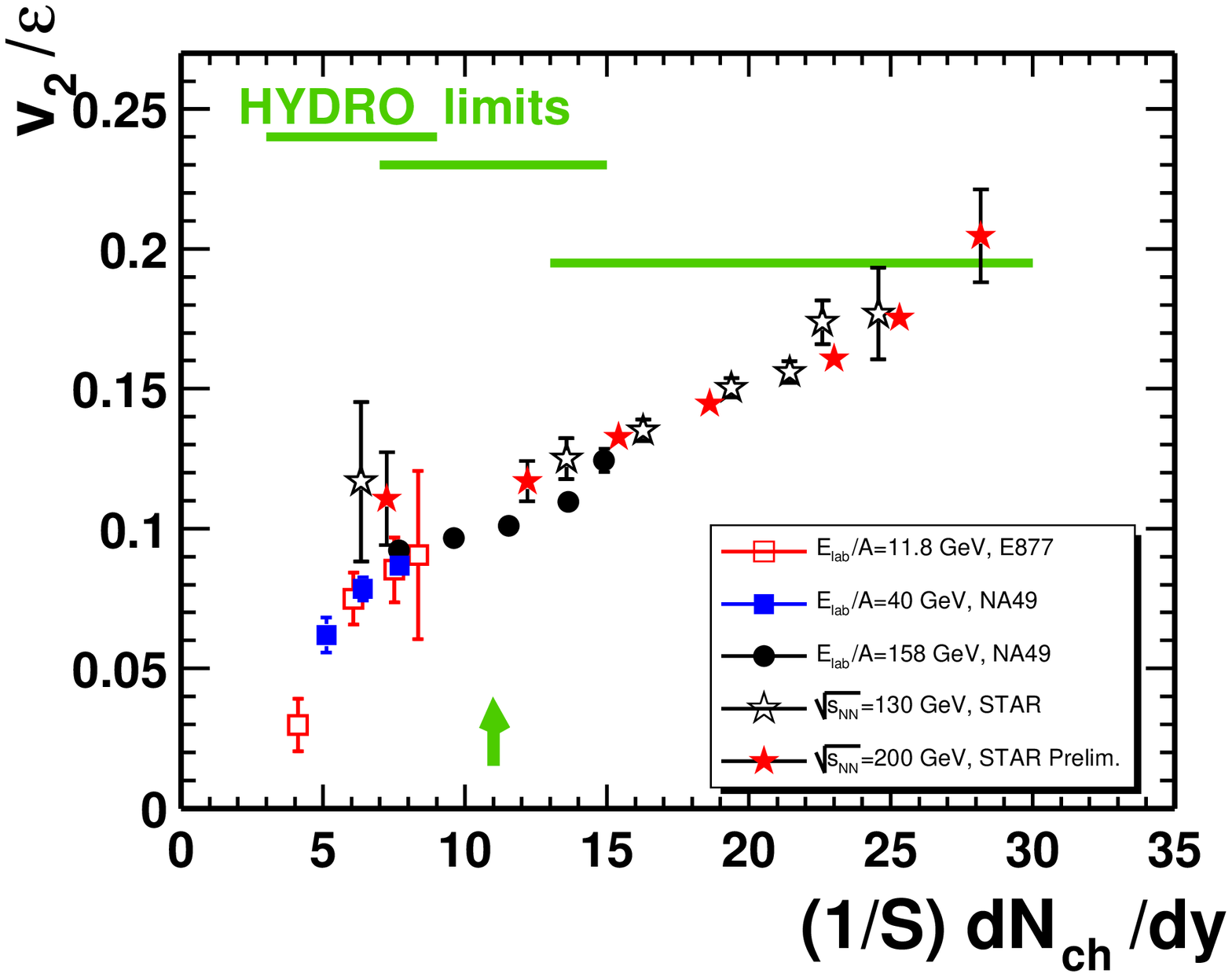}
\includegraphics[width=0.4\textwidth]{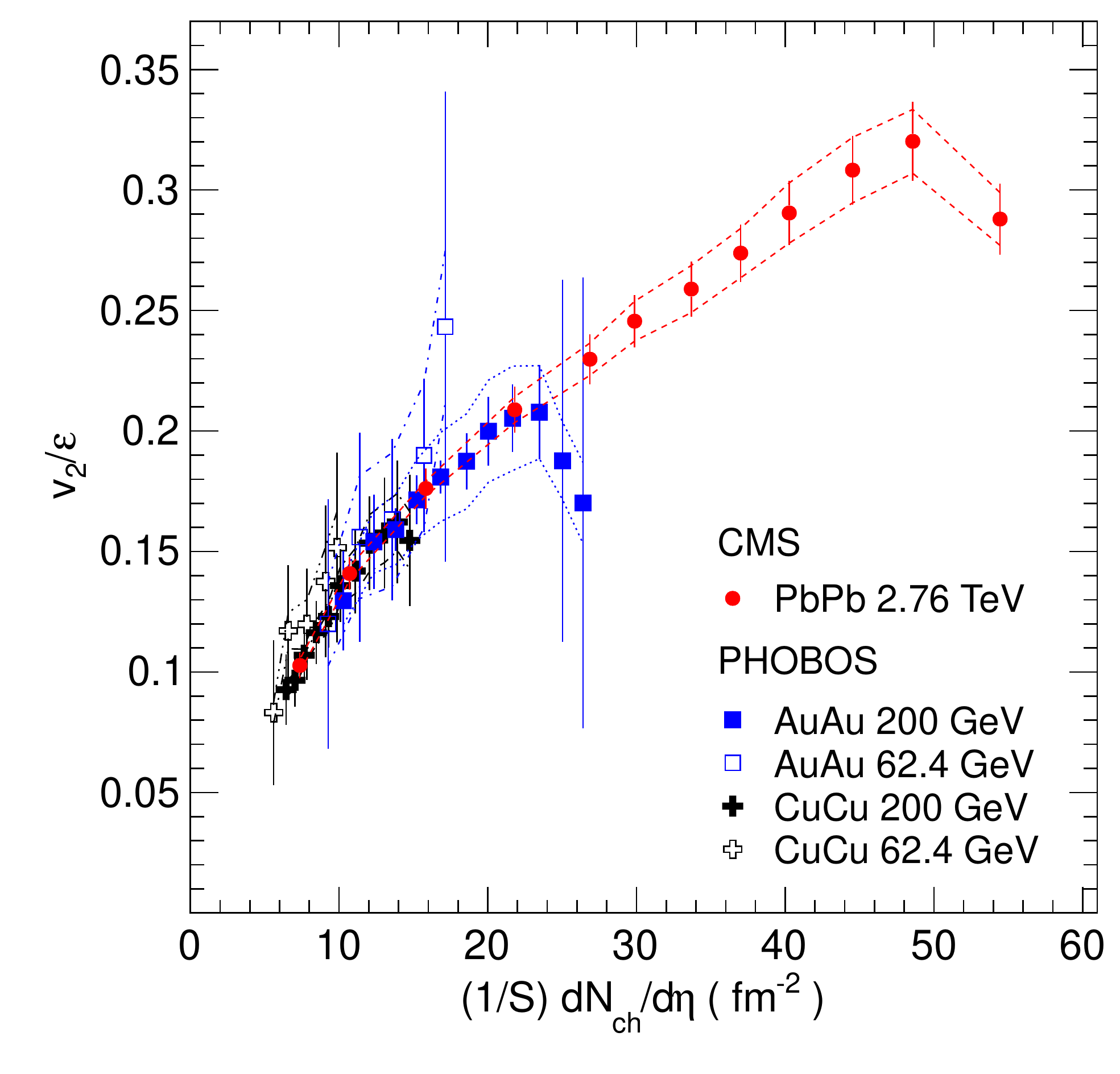}
\caption{Left panel: Elliptic flow $v_2$ scaled by spatial eccentricity $\epsilon$ as a
function of charged particle density per unit transverse area $S$,
from AGS to top RHIC energy. The hydrodynamic limit is only attained
at RHIC \cite{58}. The right panel shows the continuation of this scaled representation of the data up to Pb+Pb collisions at the LHC energy $\sqrt{s}=2.76$ TeV (from ref \cite{106}).}
\label{fig22}
\end{center}
\end{figure} \\
At top RHIC
energy, the interval between $t_0 \approx 0.5 \: fm/c$, and
hadronization time, $t_H \approx 3 \: fm/c$, is long enough to
establish dynamical consequences of an early approach toward local
equilibrium. The ''lucky coincidence'' of such a
primordial resolution of dynamical time scale, with the extreme
primordial density, offered by semi-central collisions of
heavy nuclei, results in an extremely short mean free path of the
primordial matter constituents, thus inviting a hydrodynamic
description of the expansive evolution. Thus, at RHIC energy (unlike at the SPS) the system at primordial time $t_0 \lesssim 0.5 fm/c$ is virtually "born into a hydrodynamic expansion mode". At the SPS the long interpenetration time, $t \gtrsim 1.5 \: fm/c$, confounds the primordial development and thus the hydrodynamical limit can not be reached. The system is not synchronized in its collective dynamics. 

\begin{figure}[h!]   
\begin{center}\vspace{-0.5cm}
\includegraphics[scale=1.9]{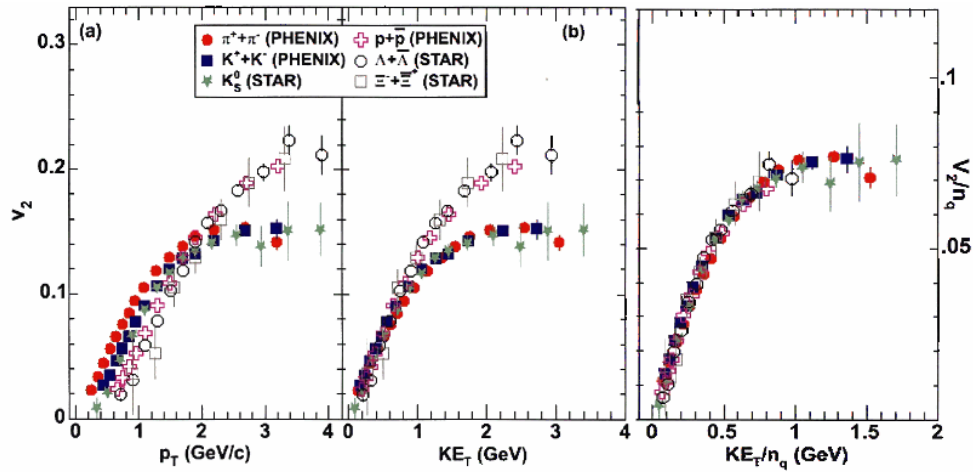}\vspace{-0.4cm}
\caption{$v_2$ vs. $p_T$ (left panel) and transverse kinetic energy
$KE_T=m_T-m_0$ (middle) for several hadronic species in min. bias
Au+Au collisions at $\sqrt{s}=200 \: GeV$, showing separate meson
and baryon branches. Scaling (right panel) is obtained by valence
quark number $n_q$, dividing $v_2$ and $KE_T$ \cite{79}.}
\label{fig23}
\end{center}
\end{figure}

\begin{figure}[h!]   
\begin{center}
\includegraphics[width=0.9\textwidth]{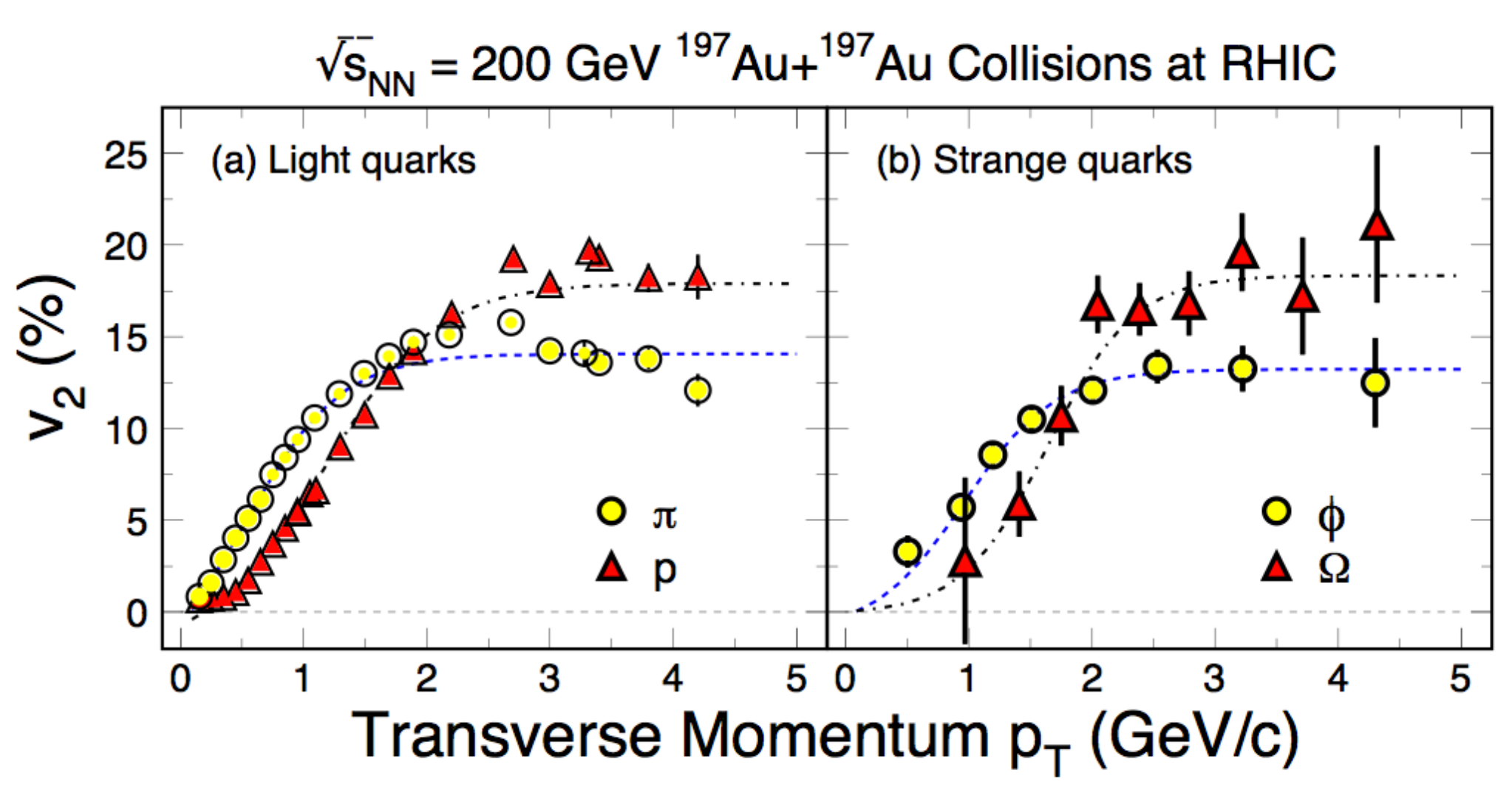}
\caption{$v_2$ vs. $p_T$ for $\pi$, $p$ (left panel) and $\Phi$ and $\Omega$ (right panel) in min. bias
Au+Au collisions at $\sqrt{s}=200 \: GeV$. The dashed lines represent fits to the meson and baryon branches, respectively \cite{80a}.}
\label{fig23a}
\end{center}
\end{figure}

A further, characteristic scaling property of elliptic
flow is derived from the $p_T$ dependence of $v_2$, observed
for the different hadronic species. In \Fref{fig21} one observes a hadron
mass dependence, the $v_2$ signal of pions and charged kaons rising
faster with $p_T$ than that of baryons. Clearly, within a
hydrodynamic flow {\it velocity} field entering hadronization,
heavier hadronic species will capture a higher $p_T$, at a given
flow velocity. However, unlike in hadronic radial expansion flow
phenomena it is not only the hadronic {\it mass} that
sets the scale for the total $p_T$ derived, per particle species,
from the elliptic flow field, but the hadronic valence quark
content. This conclusion is elaborated \cite{79} in \Fref{fig23}. \\
 \\
The left panel shows measurements of the $p_T$ dependence of $v_2$ for
several hadronic species, in minimum bias Au+Au collisions at
$\sqrt{s}=200 \: GeV$ \cite{78}. The middle panel bears out the
hydrodynamically expected \cite{80} particle mass scaling when
$v_2$ is plotted vs. the relativistic transverse kinetic energy
$KE_T \equiv m_T-m$ where $m_T=(p_T^2+m^2)^{1/2}$. For $KE_T \ge 1
\: GeV$, clear splitting into a meson branch (lower $v_2$) and a
baryon branch (higher $v_2$) occurs. However, both of these branches
show good scaling separately. The right panel shows the result
obtained after scaling both $v_2$ and $KE_T$ (i.e. the data in the
middle panel) by the constituent quark number, $n_q=2$ for mesons
and $n_q=3$ for baryons. As shown in \Fref{fig23a} this scaling is also observed for strange and multistrange baryons and mesons \cite{80a}.

The resulting perfect, universal scaling is
an indication of the inherent quark degrees of freedom in the
flowing matter as it approaches hadronization. We thus assert that
the bulk of the elliptic flow signal develops in the
pre-hadronic phase.

The above scaling analysis has also been extended to PHENIX results
\cite{81} concerning elliptic flow of the charmed $D$ meson, with perfect agreement to the observations made in
\Fref{fig23} and \Fref{fig23a}, of a separation into meson/hadron branches on the $KE_T$
scale, which merge into a universal $v_2$ scaling once both $v_2$
and $KE_T$ per valence quark are considered. The observation that
the $D$ meson charm quark apparently shares in the universal flow
pattern is remarkable as its proper relaxation time is, in
principle, lengthened by a factor $M/T$ \cite{82}. A high partonic
rescattering cross section $\sigma$ is thus required in the
primordial QGP fireball, to reduce the partonic mean free path
$\lambda=1/n \sigma$ (where $n$ is the partonic density), such that
$\lambda \ll A^{1/3}$ (the overall system size) and, simultaneously,
$\lambda \: < \: 1 \: fm$ in order to conform with the near-zero
mean free path implication of the hydrodynamic description of the
elliptic flow, which reproduces the data gathered at RHIC energy. {\it The non-perturbative quark-gluon plasma is thus a strongly
coupled state} (which has been labeled sQGP \cite{83}). At RHIC
energy this reduces the partonic mean free path to a degree that makes partonic
hydrodynamics applicable. A Navier-Stokes analysis \cite{84} of
RHIC flow data indicates that the viscosity of the QGP must be
about ten times smaller than expected if the QGP were a weakly
interacting pQCD Debye screened plasma. This justifies the use of
near-ideal fluid dynamics.

Considering first attempts to derive a quantitative estimate of the
dimensionless ratio of shear viscosity to entropy, we note that
$\eta/s$ can be estimated from the expression \cite{79}
\begin{equation}
\eta/s \approx T \lambda_f \: c_s
\label{eqn15}
\end{equation}
where $T$ is the temperature, $\lambda_f$ the mean free path, and
$c_s$ is the sound speed derived from the partonic matter EOS. A fit
by the perfect fluid ''Buda-Lund'' model \cite{85} to the scaled
$v_2$ data shown in \Fref{fig23} yields $T=165 \pm 3 \: MeV$; $c_s$ is
estimated as $0.35 \pm 0.05$ \cite{79,80}, and $\lambda_f \approx
0.30 \: fm$ taken from a parton cascade calculation including $2
\leftrightarrow 3$ scattering \cite{86}. The overall result is
\begin{equation}
\eta/s = 0.09 \pm 0.02
\label{eqn16}
\end{equation}
in agreement with former estimates of Teaney and Gavin \cite{84,87}.
This value is very small and, in fact, close to the universal lower
bound of $\eta/s =1/4\pi$ recently derived field theoretically
\cite{20}.

We take a moment here for a general remark. The pattern of the heavy ion research program has undergone a transition around about 2005, from its initial finite temperature QCD and sQGP discovery era to a second, present stage of quantitative characterization - most prominently of the collective flow observables (and of the QGP transport coefficient, in the hard sector). We have thus, in principle, finished our author's task with this article, to give a historic introduction. However, as in this case "the best seems to come at the end", we can not resist to devote a few remaining pages to the present state of the flow signals. A good part of the change to a new era did arise from the realization that, even under the most favourable conditions for the validity of the hydrodynamic description of the collisional evolution (namely at top RHIC energy with the heaviest projectiles) it became obvious that the data required some finite degree of QGP viscosity. The near-success of an ideal fluid description suggested that this viscosity had to be small. But then enter Kovtun, Son, Starinets \cite{20} and others showing that just exactly this smallness, a $\eta/s$ near $1/4\pi$, is of outstanding fundamental importance, perhaps singling out the QGP as a certain limit of all possible fluid matter \cite{90}. And, not enough with this, it turns out that this strongly coupled state should have a weakly coupled dual realization in string theory, via the AdS/CFT correspondence \cite{21}. Thus the field proceeded from "there is a small viscosity" \cite{81,82,85,87} to "how small exactly is $\eta/s$?" This has turned out to be a formidable question, and it is not finally answered until today. We note here a large number of recent, comprehensive reviews \cite{14,24,54,91,92} to which the reader is referred for detail.

\begin{figure}[h!]   
\begin{center}
\includegraphics[width=0.49\textwidth]{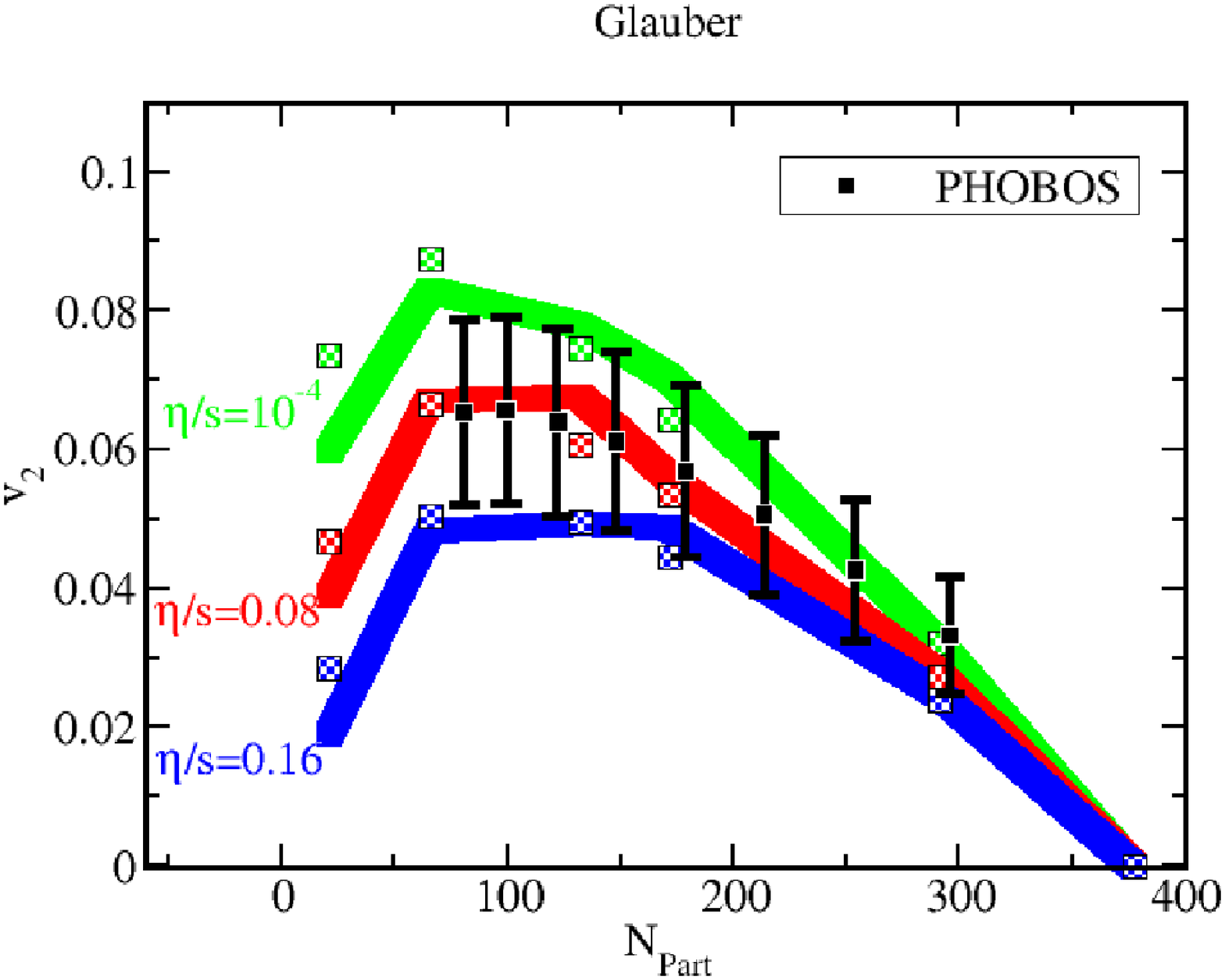}
\includegraphics[width=0.49\textwidth]{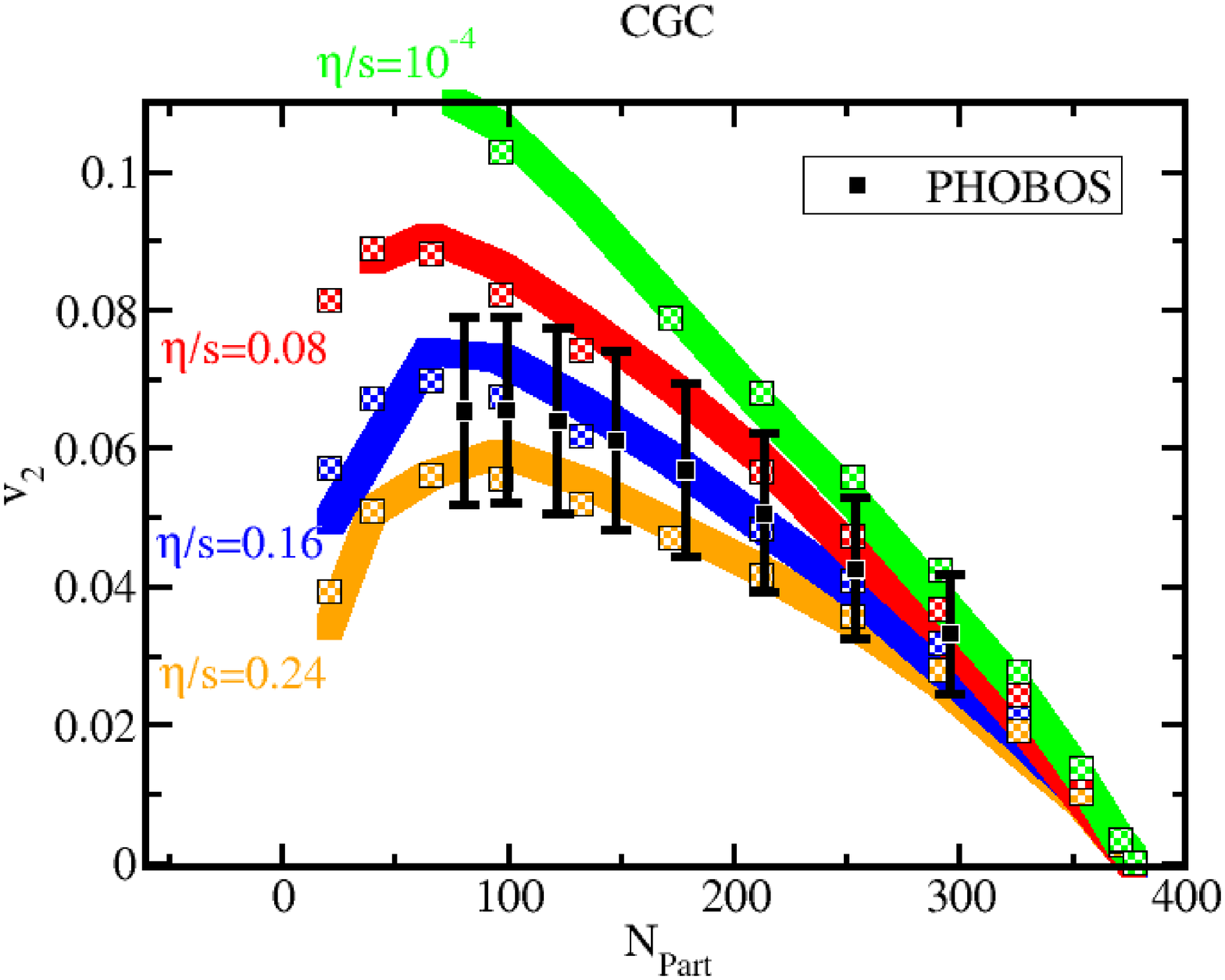}\\
\includegraphics[width=0.49\textwidth]{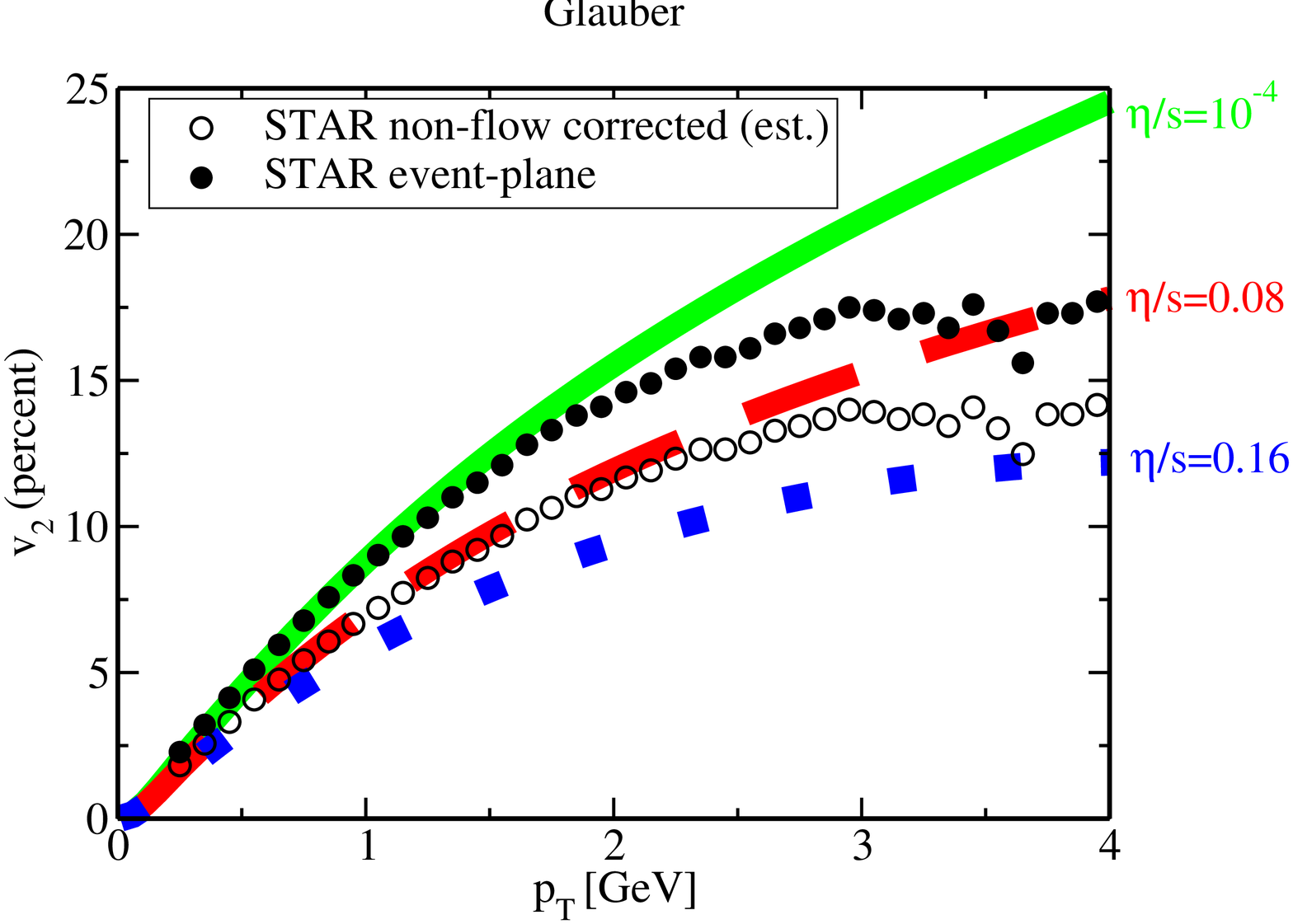}
\includegraphics[width=0.49\textwidth]{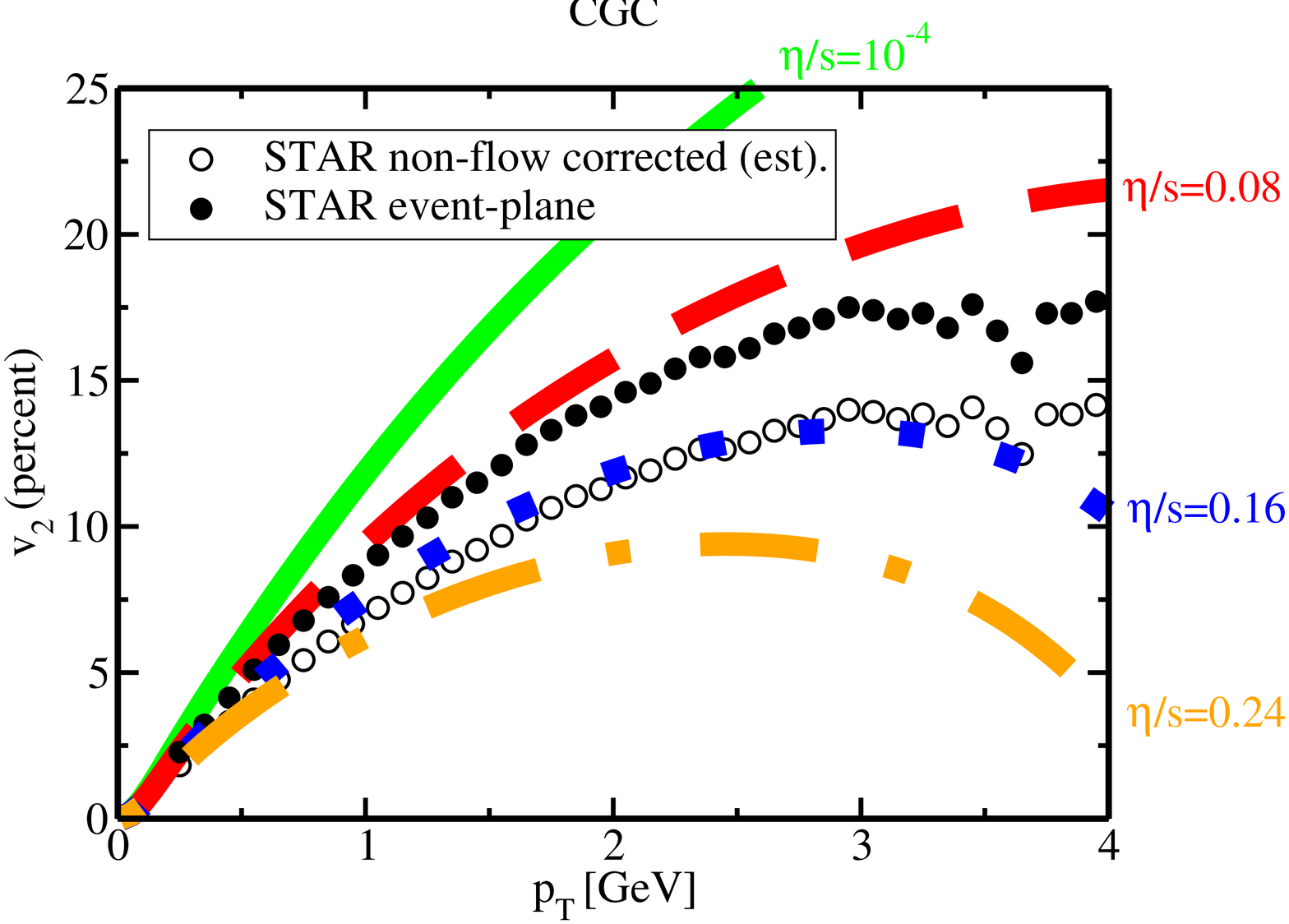}
\caption{Elliptic flow as a function of centrality (top row) and of transverse momentum(bottom row) for MC-Glauber (left column) and CGC (right column) initial conditions. Data from PHOBOS\cite{94} and STAR\cite{95} experiments for Au+Au collisions at $\sqrt{s} = 200$ GeV. The curves show viscous hydrodynamic calculations \cite{93} with specific shear viscosities $\eta/s$ as indicated. From \cite{93}.}
\label{fig24}
\end{center}
\end{figure}

"Exactly" determining the quasi-macroscopic parameters of the sQGP fluid, notably the specific shear viscosity $\eta/s$ (which controls the ability of the fireball volume to relax toward structureless equilibrium, "forgetting" the initial influences, \Fref{fig5} and \ref{fig17}), but also the sound speed, bulk viscosity, EOS etc. faces a formidable range of problems, which reside in the initialization and termination (decoupling) phases that surround the flow evolution. These phases exert their specific influences on the observed flow harmonics, interesting by themselves (existence of a CGC state of QCD) but counterproductive as far as exact determination of $\eta/s$ is concerned. A vivid illustration of such an interplay of influences is illustrated in \Fref{fig24} \cite{93}. It shows elliptic flow $v_2$ as a function of centrality(top row) and of transverse momentum(bottom row), for viscous hydro calculations with Glauber-MC(left column) and Colour Glass Condensate(CGC) initialization. The experimental data in the top row are from PHOBOS for top RHIC energy \cite{94}, in the bottom we see STAR results \cite{95} in two different versions, one from straight forward event plane analysis (full points), the other after correction for background, non-flow effects (open points). All data refer to Au+Au collisions at mid-rapidity. The $v_2$ coefficient is seen to decrease from peripheral to central collisions, along with the decreasing initial spatial excentricity from \Eref{eqn13}. And to increase towards high $p_T$ as harder hadrons offer a higher spatial resolution of the initial impact configuration (\Fref{fig4}). The curves show viscous hydro calculations \cite{93} of charged hadron elliptic flow, with Glauber and CGC initializations, for several values of $\eta/s$ as indicated. These calculations employ an EOS from lattice QCD \cite{96} above the hadronization temperature, $T_c = 165$ MeV, matched at $T_c$ to the EOS of a chemically equilibrated hadron fluid at lower temperatures. The latter choice is debatable \cite{69,97} as we shall see below. Anyhow: we see that the fits with Glauber-MC initialization clearly indicate a value $\eta/s$ = 0.08 (with the corrected STAR data), wheras the fits with initial CGC prefer the value 0.16, with quite some sensitivity. A factor of two.

A short reminder. Glauber-MC initialization \cite{98} translates the initial target-projectile Woods-Saxon distribution of nucleons, via straight Glauber trajectories, into a smeared collision point density in the transverse plane that is interpreted as the primordial energy density distribution from which the hydro evolution originates. The CGC model envisions the initial creation of a novel QCD state, governed by quasi-classical fields of spatially unresolved gluons \cite{23}. It decays after about 0.5 fm/$c$ into the partonic transverse density distribution, i.e. into the hydro startup state. It turns out that the latter density distribution is narrower (more focused) than the Glauber version, thus explaining the higher elliptic flow from CGC initialization. Facing the data this hydro fit thus requires a higher degree of attenuation by viscous "blurring" of the initial imprint (\Fref{fig4}), i.e. a larger $\eta/s$, in agreement with the conclusion from \Fref{fig24}.

We note that two further initialization models are widely employed. The straight trajectories of nucleons in the Glauber-MC approach can be replaced by a full fledged microscopic transport model like UrQMD \cite{99} which runs until the end of the interpenetration phase, then to be translated into the hydro startup density distribution. The participant nucleon number, and the initial excentricity of the participants are an event by event side result. Also the CGC initialization has a closely related variant, the so-called MC-KLN model \cite{100}. It samples the initial nucleons as in the Glauber approach but proceeds to calculate the entropy or energy density as created by the initial creation of gluons, calculated by merging of two gluons from the projectile and target nuclei, respectively, from gluon structure functions which implement the gluon saturation effects \cite{24} also envisioned in CGC theory.

Thus there are (at least) four initialization methods. This plurality of approaches goes on with the hydro flow phase. The method of choice \cite{92} is second order viscous relativistic fluid dynamics, introduced by Israel and Stewart \cite{101} in 1979 and recently explored in heavy ion collision physics \cite{70,102}. For an introduction to these complicated matters the reader is referred to refs \cite{14,92}. A particular adaptation of this theory \cite{93} was employed in \Fref{fig24}. Several alternative effective theories are in use that are beyond our introductory purposes, but all viscous hydro models universally reduce all flow coefficients $v_n$, in comparison to ideal hydrodynamics. \Fref{fig24} shows a strong sensitivity to the choice of $\eta/s$ for elliptic flow $v_2$.

At top RHIC and LHC energies the hydro flow phase exerts the most prominent influences on the flow signals $v_n$, in comparison to the other key ingredients, the initialization and the decoupling phases. We have not yet addressed the latter, final part of the dynamical evolution. Let us note that the simplest approach, to run a hydro description until the final decoupling (the so-called "kinetic freeze-out'' from strong interaction, at T = 100 MeV)
has turned out to be inadequate \cite{69}. The hadron/resonance expansion phase that follows hadronization is characterized by a viscosity much higher than that of the preceeding QGP phase, thus its influence on flow signals must be addressed carefully. In particular, the mean free path increases well beyond 1 fm, and the hadronic population deviates far from thermal equilibrium because it freezes out already at, or near the QCD hadronization temperature, T = 160 MeV \cite{42,96}. The hydro evolution should thus be terminated at hadronization, and attached to a microscopic hadron transport model like UrQMD \cite{50,97,99}, via the Cooper-Frye prescription \cite{103}. The transport model acts as a so-called "afterburner".

\begin{figure}[h!]   
\begin{center}
\includegraphics[width=0.9\textwidth]{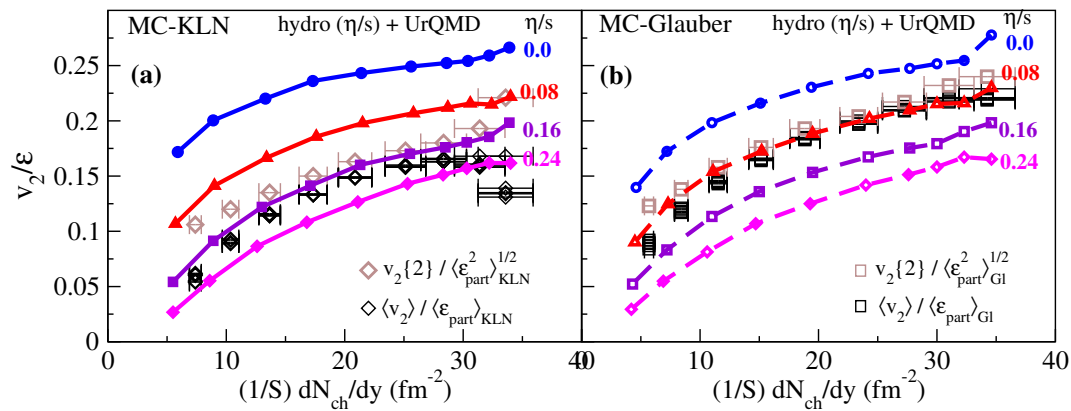}
\caption{The excentricity-scaled elliptic flow of charged hadrons, as a function of total charged hadron multiplicity density per unit overlap area. The data points show two measures of elliptic flow, from Au+Au minimum bias collisions at top RHIC energy \cite{104}. The theoretical curves \cite{92} show different choices of QGP specific shear viscosity. From \cite{92}.}
\label{fig25}
\end{center}
\end{figure}

\Fref{fig25} shows a recent analysis \cite{92} of experimental $v_2$ signals from top RHIC energy Au+Au minimum bias collisions \cite{104} in which all the above requirements are in place: state of the art in 2012. It follows the idea, illustrated in \Fref{fig22}, to employ a scaled representation \cite{72}: elliptic flow $v_2$ relative to the initial (impact parameter dependent) average spatial excentricity $\epsilon_2$ of the primordial fireball source, as a function of total charged particle mid-rapidity density per unit target-projectile overlap area S. The STAR data points \cite{104} result from two, closely related methods of determining $v_2$ (see \cite{104} for detail).
The left panel curves illustrate the predictions employing the MC-KLN initialization, the right panel uses the MC-Glauber model. Drawn for successive choices of $\eta/s$, we see a fit preference for $\eta/s$ in the vicinity of about 0.2 for the former model, and for about 0.1 with the latter.

Essentially no progress toward precision in $\eta/s$ resolution, in comparison to the 2008 analysis shown in \Fref{fig24}. The influences from initialization, and from the hydro evolution are intertwined in the elliptic flow variable. We see here an example of the mapping of initial geometry, and energy density input by the hydro flow field: it delivers at the end what was imprinted initially. The next task is, therefore, to find other observables that could bring selective evidence for the validity of
either one of the initialization models. In the last section we shall turn to such attempts.

\section{Higher Harmonics and Event-by-Event Analysis}\label{SectV}

In this last chapter we shall give a sketchy, mostly narrative description of the theoretical concepts, and experimental results recently obtained in attempts to progress beyond the insight obtained from comprehensive elliptic flow analysis. The latter has presented evidence for the presence of two major influences on the observed anisotropic flow:
the initialization conditions, determined by the primordial physics governing the interval $t < 0.3$ to $0.5 fm/c$, and the subsequent QGP evolution described by viscous hydrodynamics. Both these influences are of fundamental interest. Can we find evidence for, or against the existence of a primordial Colour Glass Condensate(CGC) state of QCD, perhaps even delineating some of its properties, from sharpening the focus on its outcome, i.e. the higher moments of the initial excentricity that initialize the hydro phase at about $t = 0.5 fm/c$? And, second, can we decouple, from the initial conditions, the second fundamental influence, exerted by the "near minimal" shear viscosity $\eta/s$, a crucial collective property of sQGP?

In response to these two questions of key importance, an unprecedented effort is being made by (up to 30) theory groups elaborating selective modes of observation, far beyond the "simple" $v_n$ harmonics of \Eref{eqn1},
the analysis of which was actually pioneered long ago, by the E877 experiment at the AGS \cite{105}. These recent developments are described in references \cite{54,91,92}, and we shall illustrate some of them below. Moreover, the recent era of high statistics Au+Au and Pb+Pb runs at RHIC and LHC has enabled the PHENIX, PHOBOS, STAR, ALICE, ATLAS and CMS collaborations to come up with precision measurements of virtually all relevant observables, outlined so far. High statistics also invites a novel analysis idea, to focus on "super-central" events \cite{106}, from the maximum multiplicity end range of charged hadron multiplicity distributions, which should be most selective to initialization models, and less so to impact geometry effects. Furthermore, the high multiplicity of final hadrons, of order 2000 per unit rapidity at the LHC energy of $\sqrt{s} = 2.76$ TeV, permits two new forms of event by event analysis. Firstly the so-called "event engineering" \cite{107} analysis where each event is divided up into two subevents. The first gets analyzed, for its $v_2$ or $v_3$ flow, and a sample of maximum or minimum flow events can then be analyzed, in its second half, for the same, or different observables, assessing its correlation with extremal $v_n$. The second new event by event analysis technique attempts to "unfold" \cite{108} the true eventwise $v_n$ distribution from the influences of finite number fluctuations and non-flow effects. These distributions should reveal particular properties (lumpiness, fluctuations) of the primordial source, helping to distinguish CGC from trivial Glauber-MC initial states.

\begin{figure}[h!]
\begin{center}
\includegraphics[width=0.33\textwidth]{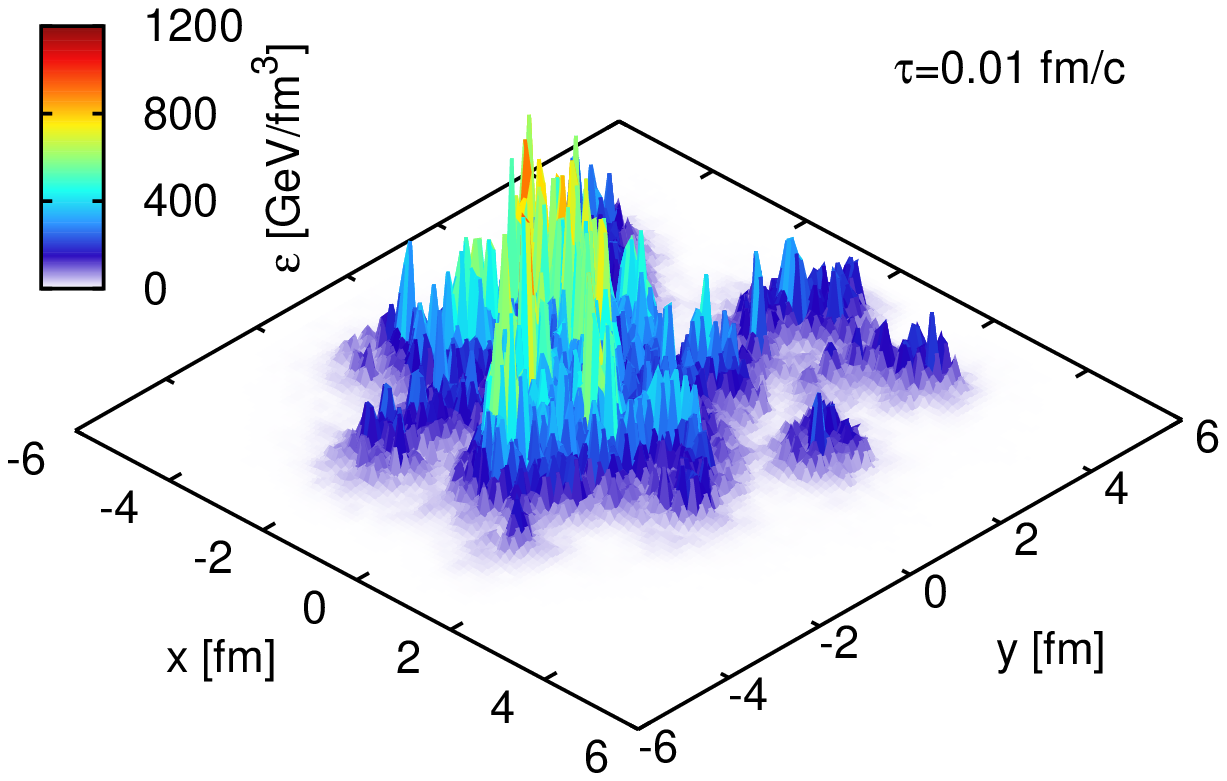}
\includegraphics[width=0.33\textwidth]{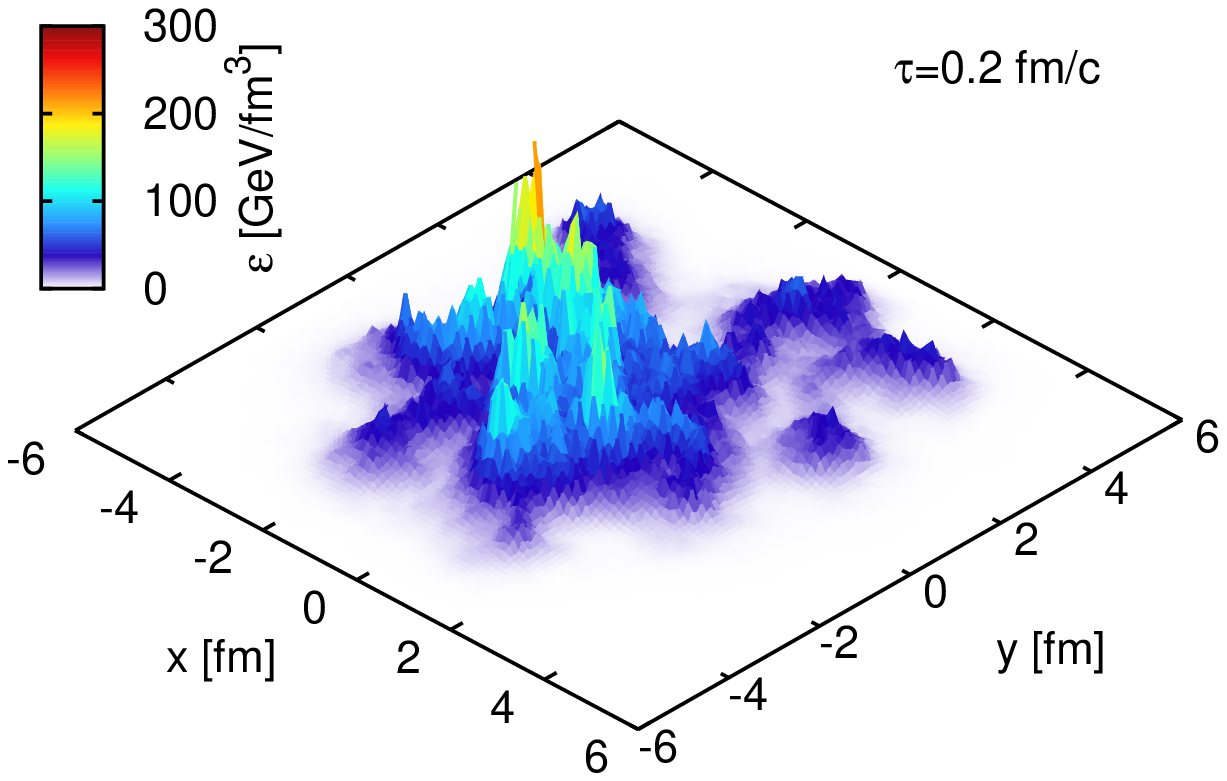}
\includegraphics[width=0.33\textwidth]{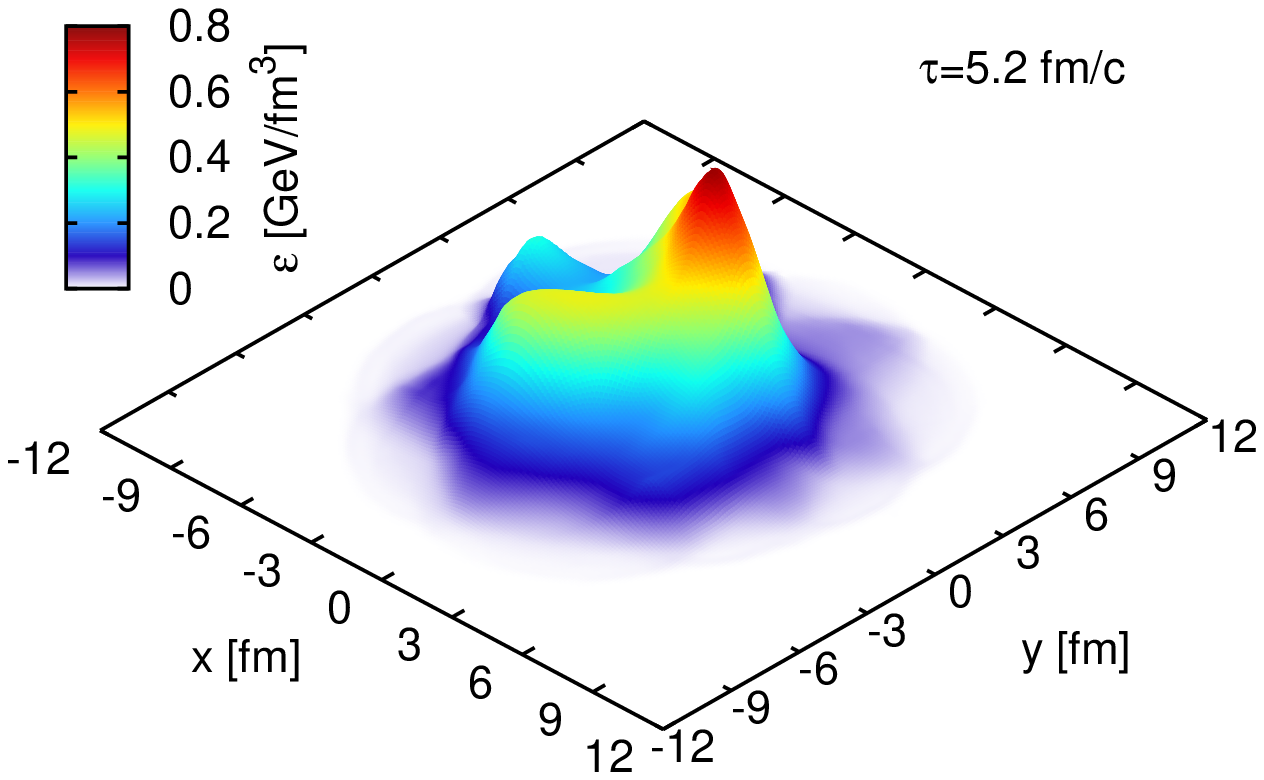}
\caption{The transverse energy density profiles from the CGC model for a
 semiperipheral Au+Au collision at $\sqrt{s}=200$ GeV, at times
 $t = 0.01, 0.2$ and $5.2 fm/c$, showing smoothening from the initial
 CGC ``glasma'' state, via hydro-initialization to freeze-out
 conditions, with viscous hydrodynamic evolution employing
 $\eta/s = 0.12$ (from ref.\cite{92}).}
\label{fig26}
\end{center}
\end{figure}

We begin a brief illustration returning to the initialization of a single event in \Fref{fig26} \cite{92}. It shows the transverse energy density distribution from CGC theory \cite{109} for a semi-peripheral Au+Au collision at $\sqrt{s}=200$ GeV, at times $t = 0.01, 0.2$ and 5.2 $fm/c$. The influence of impact geometry is hardly recognizeable initially, from among the dramatic, out of equilibrium fluctuation and lumpiness. At $t = 0.2 fm/c$, there is already some smoothing, and the energy-momentum tensor T(x,y) from the CGC evolution is matched to the form used in the hydro model; a viscous hydro flow phase follows assuming $\eta/s$ = 0.12. This smoothes the fluctuations to some degree (proportional to $\eta/s$) but some memory remains. It is the purpose of event-by event analysis, both theoretically and experimentally, to identify and quantify the model-specific fluctuations that would be averaged out in an ensemble analysis.

Theory has preferred to stick with the spherical harmonics Fourier analysis, augmented by many more complicated quantities (mixed harmonics, cumulants, differential or $p_T$ and/or $y$ averaged) which we can not illustrate here (see,e.g. ref. \cite{54} for review). Such observables describe the event density distribution "as seen from the outside" (by the emitted flow), which might not be the last word if one wants to characterize internal hot spots and valleys.

\begin{figure}[h!]
\begin{center}
\includegraphics[width=0.5\textwidth]{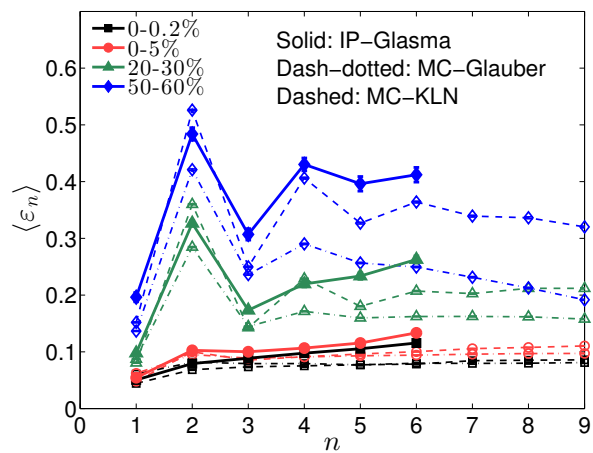}
\caption{Primordial fluctuation Fourier harmonics of energy density
 excentricities $\epsilon_n$ in Pb+Pb collisions at 2.76 TeV
 for the initialization models CGC-Glasma, MC-Glauber and MC-KLN,
 as a function of event centrality starting with ``supercentral''
 collisions. From ref \cite{92}.}
\label{fig27}
\end{center}
\end{figure}

\Fref{fig27} shows an example \cite{92} of attempts to describe the influence of different initialization models, focusing on the initial excentricity moments $\langle\epsilon_n\rangle$, equivalent to \Eref{eqn13}, of Pb+Pb collisions at 2.76TeV, for four centralities and CGC, MC-Glauber and MC-KLN initialization.

Without fluctuations the odd moments would be zero, and
$\langle\epsilon_n\rangle$ would approach zero in zero impact parameter collisions. A drastically different situation is seen here: for the "super-central" event selection, all excentricity moments are finite but essentially flat, thus indicating a spherical, slightly oscillating source. This centrality cut thus selects for non-geometry related flow from the fluctuations. Proceeding to less central collisions, the nuclear overlap region develops the pronounced geometrical pattern of \Fref{fig4}, elliptic geometric deformation of the source which increases $\langle\epsilon_2\rangle$, $\langle\epsilon_4\rangle$ and $\langle\epsilon_6\rangle$ whereas, in particular, $\langle\epsilon_3\rangle$ remains essentially fluctuation dominated \cite{92}. Thus, to some degree $v_2$ and $v_3$ have complementary informational content \cite{110}. It is concluded that systematic, high precision studies of a wide range of flow harmonics could help to constrain the initial state models \cite{92}, ideally enabling one to decide on the correct one. This beeing settled one could focus on the shear viscosity $\eta/s$ \cite{111}, by $v_2$ vs. $v_3$ analysis \cite{112}.

\begin{figure}[h!]
\begin{center}
\includegraphics[width=0.9\textwidth]{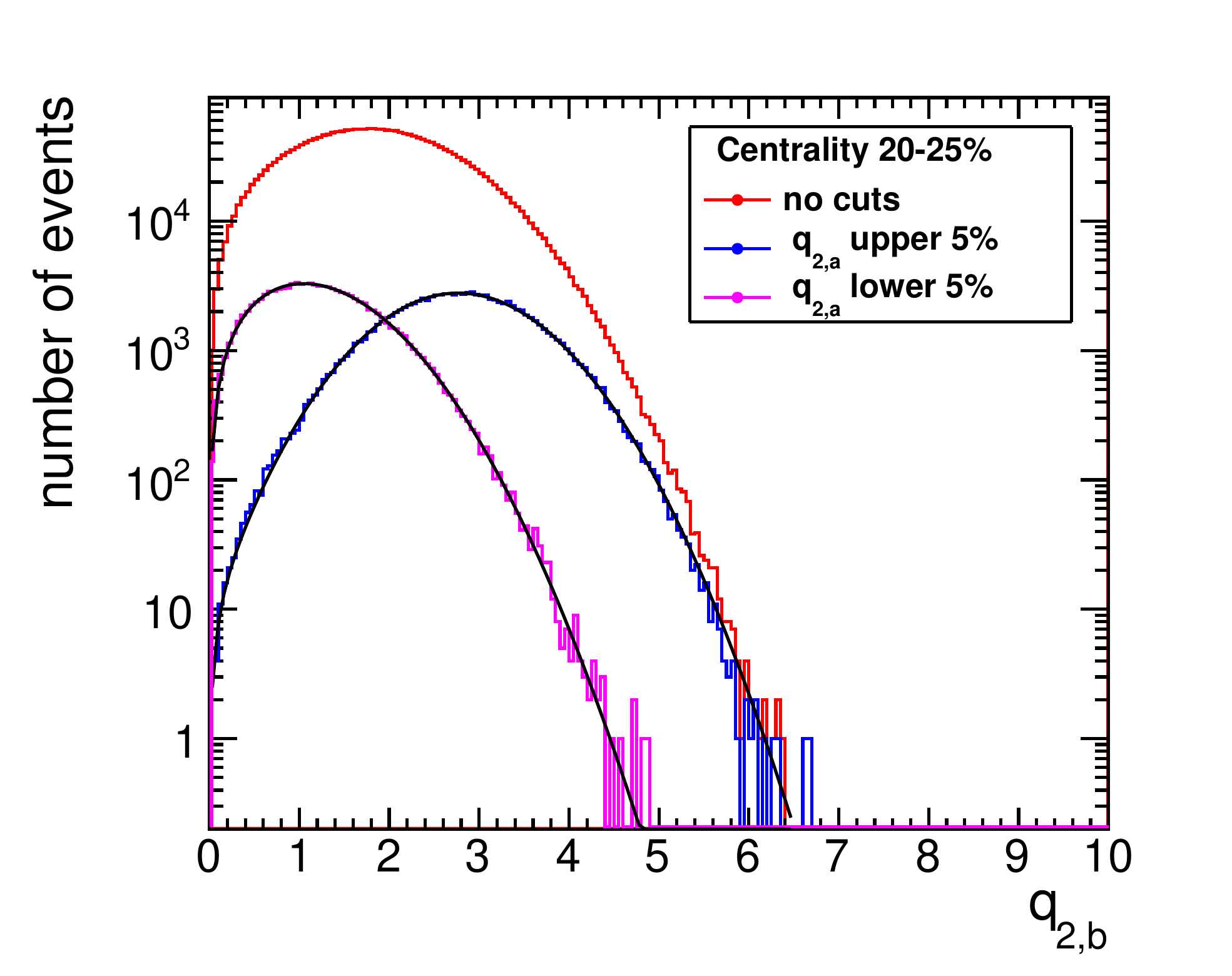}
\caption{Illustration of the selection power from event engineering: the
 distributions of the eventwise elliptic flow vector in the ``b''
 subevents after engineering cuts on minimal and maximal elliptic
 flow in the ``a'' subevents (from ref \cite{107}.)}
\label{fig28}
\end{center}
\end{figure}

Turning to the new forms of event by event analysis of flow variables, we illustrate "event engineering" in \Fref{fig28}, from a MC model study \cite{107} that generates simulated events roughly resembling LHC multiplicity conditions.

The eventwise $v_2$ and $v_3$ values are known by construction. Their distributions are then convoluted with finite number statistical fluctuations and non-flow fluctuations, to simulate the observational conditions. From these "realistic" events two subevents "a" and "b" are formed. For the a-subevents an "experimental" flow analysis is performed employing the flow vector method (already mentioned in \Sref{SectIII}, \Eref{eqn7},(\ref{eqn8})) by determining the 2-dimensional transverse event ``flow vector'' with components

\begin{equation}
Q(n,x)= \sum_i^M \cos(n \times \phi_i)\\
Q(n,y)=\sum_i^M \sin(n \times \phi_i).
\label{eqn17}
\end{equation}

The sum over the azimuthal angles $\phi_i$ of particles $i$, up to M the multiplicity of the subevent a, represents a random walk which, in the absence of correlations, would have a total lenght proportional to $\sqrt{M}$. A normalized flow vector is then defined for each a subevent,
\begin{equation}
\vec{q_n} = \vec{Q_n}/\sqrt{M}.
\label{eqn18}
\end{equation}

The idea of this investigation is to demonstrate the selective power of "engineering" the distribution of $|q_n|$ in the a-subevent ensemble, via cuts on the upper and lower 5\% of this distribution (maximal and minimal flow events, respectively). \Fref{fig28} shows the $|q_2|$ distributions then observed in the corresponding b-subevent ensembles. Indeed, the distribution for the "high $v_2$ flow engineering" reveals a much higher flow average (and width). In a similar analysis, with real events, one would thus accomplish a selection according to maximal, or minimal $v_n$ flow event classes, which would then be analyzed separeately with the hydro model. We should expect the former event ensemble to react more sensitively to the choice of specific shear viscosity $\eta/s$, the latter to be more sensitive to the initialization model.

\begin{figure}[h!]
\begin{center}
\includegraphics[width=0.36\textwidth]{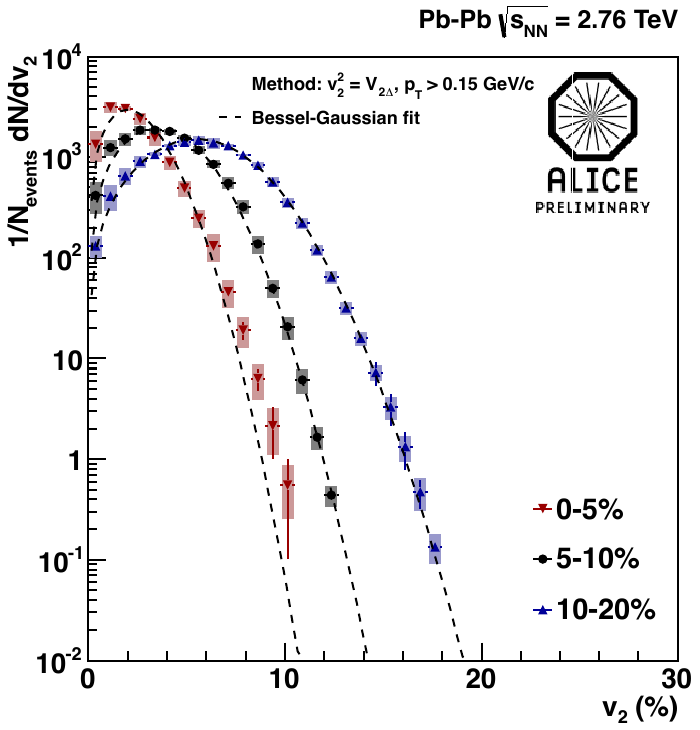}
\includegraphics[width=0.63\textwidth]{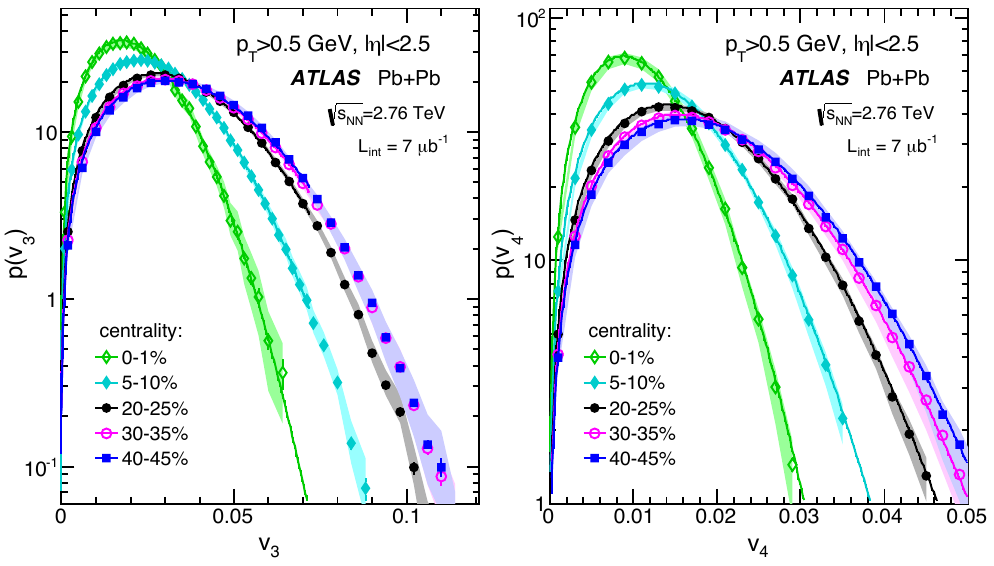}
\caption{Unfolded event by event distributions of the flow harmonics $v_2$,
 $v_3$ and $v_4$, for successive centralities, in Pb+Pb collisions at
 2.76 TeV, from the ALICE and ATLAS experiments \cite{108}. Figure from
 ref \cite{54}.}
\label{fig29}
\end{center}
\end{figure}

A final, novel technique employed in event by event analysis is based on a related, but inverse approach. In the above model study the "true" values of the flow harmonics $v_n$ are known for each event, by construction, then to be convoluted with the finite number etc. fluctuation effects besetting the experimental observation. The inverse approach attempts to de-convolute for such influences the observed event by event distributions of the flow harmonics, to arrive at the "true" distributions of $v_n$ \cite{108}. Recalling that the shape, and width of these true distributions closely reflect the fluctuations of the primordial transverse energy density profiles as illustrated in \Fref{fig5} and \ref{fig26}, this method offers a novel chance of discriminating among the initial dynamics models. Can we find evidence for, or against the CGC "glasma" state of QCD? We show in \Fref{fig29} the results \cite{108} of this unfolding procedure, based on event by event $v_2$, $v_3$ and $v_4$ flow analysis by the ALICE and ATLAS LHC experiments, for Pb+Pb at $\sqrt{s}=2.76$ TeV, analyzed in successive bins of centrality. The mean, and widths of the "true" distributions shift upward toward semi-peripheral collisions; for the mean values this shift reflects the increasing influence of geometry on the primordial excentricity modes, as illustrated in \Fref{fig27}. The shape and width details will present considerable challenges to theoretical analysis.

Our "historical introduction" thus ends with this 2014 perspective. It leads us to expect that the dramatic recent expansion of theoretically formulated collective flow observables, and of their swift coverage by high resolution experimental analysis, will finally find the answer to the main question of "exactly what" are the collective properties of deconfined QCD matter.

We would like to thank Julian Book for his invaluable contributions to realizing this manuscript. We thank Art Poskanzer and Alexander Schmah for valuable discussions and suggestions, and our referee from journal J.Phys.G for a critical reading, resulting in substantial revision. This work was supported in part by Deutsche Forschungsgemeinschaft (DFG), HIC for Fair and by the office of NP within the U.S. DOE Office of Science.

\end{document}